\documentstyle[11pt,aaspp4]{article}

\begin{document}

\received{}
\revised{}
\accepted{}

\lefthead{Catelan et al.}
\righthead{Bimodality and Gaps on Globular Cluster Horizontal Branches}

\slugcomment{To appear in The Astrophysical Journal}

\singlespace

\title{Bimodality and Gaps on Globular Cluster Horizontal Branches.\\ 
II. The Cases of NGC 6229, NGC 1851 and NGC 2808}

\author{M.~Catelan \altaffilmark{1} }
\authoraddr{catelan@stars.gsfc.nasa.gov}
\author{J.~Borissova \altaffilmark{2} }
\authoraddr{jura@hmm.bg400.bg}
\author{A.~V.~Sweigart \altaffilmark{1} }
\authoraddr{sweigart@bach.gsfc.nasa.gov}
\and
\author{N.~Spassova \altaffilmark{2} }
\authoraddr{neda@center.phys.acad.bg}

\altaffiltext{1}{
   NASA/Goddard Space Flight Center, Code 681, Greenbelt,
   MD 20771, USA
   }
\altaffiltext{2}{
   Institute of Astronomy, Bulgarian Academy of Sciences,
   72~Tsarigradsko chauss\`ee, BG\,--\,1784 Sofia, Bulgaria
   }

\begin{abstract}
The outer-halo globular cluster NGC 6229 has a
peculiar horizontal-branch (HB) morphology, with clear indications
of a bimodal HB and a ``gap" on the blue HB. In this
paper, we present extensive synthetic HB simulations
to determine whether
peculiar distributions in the underlying physical parameters are
needed to explain the observed HB morphology. We find that
a unimodal mass distribution along the HB can satisfactorily
account for the observed HB bimodality, {\em provided} the mass dispersion
is substantially larger than usually inferred for the Galactic globular
clusters. In this case, NGC 6229 should have a well-populated, 
extended blue tail. A truly bimodal
distribution in HB masses can also satisfactorily account for the
observed HB morphology, although in this case the existence of an 
extended blue tail is not necessarily implied. The other
two well-known bimodal-HB clusters, NGC 1851 and NGC 2808, are
briefly analyzed. While the HB morphology of NGC 1851 can also be 
reproduced 
with a unimodal mass distribution assuming a large mass dispersion, 
the same is not true of NGC 2808, for which a bimodal, 
and possibly multimodal, mass distribution seems definitely required. 

The problem of gaps on the blue HB is also discussed. Applying the
standard
Hawarden (1971) and Newell (1973) $\chi^2$ test, we find that 
the NGC 6229 gap is significant at the $99.7 \%$ level.
However, in a set of 1,000 simulations, 
blue-HB gaps comparable to the observed one are present
in $\sim 6 \% - 9 \%$ of all cases. We employ a 
new and simple formalism,
based on the binomial distribution, to explain the origin of this 
discrepancy, and conclude that Hawarden's method, in general, 
substantially overestimates the statistical significance of gaps.

\end{abstract}


\clearpage

\section{Introduction}
The outer-halo globular cluster (GC) NGC 6229 (C1645+476) has recently
been studied photometrically by Borissova et al. (1997, hereafter Paper I). 
These authors showed that the horizontal-branch (HB) morphology of this
cluster is characterized by two peculiar properties: i) A distinctly
bimodal distribution in $\bv$ colors, consistent with the relatively
small number of RR Lyrae variables, compared to the
blue HB and red HB populations; and ii) At least one
``gap" on the blue HB, as previously 
noted by Carney, Fullton, \& Trammell (1991).

Bimodal-HB clusters are relatively rare, as are clusters with gaps 
on the blue HB. NGC 6229 is one of only two known cases---the other
being NGC 2808 (Sosin et al. 1997a)---where these two anomalies are 
simultaneously present. As argued by many authors,
an understanding of the nature of the detected peculiarities in 
``bimodal" and ``gap" clusters would be of paramount importance
for understanding the nature of the second parameter phenomenon
(e.g., Buonanno, Corsi, \& Fusi Pecci 1985; Rood et al. 1993; Stetson, 
VandenBerg, \& Bolte 1996; Paper I). 

These considerations have 
prompted us to perform a theoretical investigation 
of the problem of HB bimodality and gaps. In particular, we have considered 
the following questions: i) Can HB bimodality be produced by canonical
evolutionary models? ii) If so,
what are the requirements for producing HB bimodality 
in the specific cases of NGC 6229, NGC 1851, and NGC 2808? 
iii) Are gaps on the blue HB a natural consequence of stellar evolution, 
as claimed by some authors? iv) What is the statistical significance 
of gaps?

In this paper, we present extensive synthetic HB simulations
for NGC 6229. We begin in the next section by laying out a proposed 
nomenclature for both defining and differentiating ``bimodal" and 
``gap" clusters. In Sec. 3, we describe the theoretical procedure used 
in Sec. 4 to compute synthetic HBs for NGC 6229. The statistical significance 
of gaps on the blue HB is next discussed in Sec. 5. In Sec. 6, we present 
additional computations for the bimodal-HB clusters NGC 1851 and NGC 2808.
In Sec. 7, we discuss some previously proposed 
scenarios for the origin of HB bimodality and gaps, in the light of
the present results. Finally, Sec. 8 summarizes our results.
In an Appendix, a compilation is presented and discussed of GCs which
may present signs of HB bimodality and gaps, according to the definitions
laid out in Sect. 2. Several of these clusters may require more detailed
observational and theoretical investigations in the future.

\section{Definition of HB Bimodality and Gaps}

Before discussing HB bimodality and gaps, it is important to clearly 
define these terms, since they have been used somewhat ambiguously
in the literature. This is especially important for bimodality, because
the presence or lack of ``bimodality" depends on the quantity being
considered. For instance, a cluster may be bimodal in $\bv$, but not in
$T_{\rm eff}$ or mass.

The term ``bimodal HB" has sometimes been applied to
clusters with a gap on the blue HB (e.g., Crocker, Rood, \& O'Connell 1988). 
More traditionally (e.g., Norris 
1981; Lee, Demarque, \& Zinn 1988), this term has been used to 
characterize those clusters having two {\em main modes} in the $\bv$ 
color distribution, one on the red HB and one on the blue HB. 

We shall call {\em ``bimodal-HB clusters"} those objects which
show a dip in the {\em number counts} at the RR Lyrae level, in comparison
with both the blue HB and the red HB. In other words, a bimodal-HB
cluster, in this scheme, is one for which $B > V < R$, where
$B$, $V$, and $R$ are the number of stars on the blue HB, 
inside the instability strip, and on the red HB, respectively.
Such a definition will generally imply that the distribution
in $\bv$ colors along the HB has two preferred statistical 
modes as well---one on the red HB clump, and one on the blue HB.  

We emphasize that this definition of HB
bimodality does not necessarily imply that {\em all} quantities
characterizing an HB distribution will have two preferred
statistical modes. For instance,
the distribution in magnitudes may very
well have just one mode, as is commonly the case.
Most importantly, whether the underlying {\em temperature} 
and/or {\em mass} distributions 
have two preferred statistical modes as well, as we shall see, 
can only be determined after a thorough case-by-case study. 
The main advantage of our definition is that it is based entirely
on {\em directly observed quantities}; its purpose is to
serve as a helpful guide when searching for clusters with 
potentially more than one statistical 
mode in the underlying physical parameters, especially the 
temperature and mass. At any rate (see also Rood \& Crocker 
1985b), our definiton of HB bimodality immediately suggests
an important difference between the bimodal-HB clusters 
thus defined and the archetypal ``standard" cluster M3 = NGC 5272, 
for which $B: V: R \simeq 0.33: 0.41: 0.26$ (Buonanno et al. 1994). 

We shall call {\em ``gap clusters"} those objects which show
more or less pronounced gaps along the blue HB, or possibly
even on the red HB or {\em inside} the instability strip. 
In those instances where a gap {\em coincides} with the 
instability strip---as in the case of NGC 2808 (Harris 1974)---we 
shall call the object a ``bimodal-HB cluster," {\em not} a ``gap 
cluster." A bimodal-HB cluster may, however, simultaneously
be classified as a gap cluster, provided an additional gap is
present somewhere along the blue or red HB. It is interesting 
to note that only four objects have thus far been 
simultaneously classified as ``bimodal-HB" and 
``gap" clusters, namely: NGC 2808, NGC 6229, NGC 6388, and 
NGC 6441.

In the Appendix, we present the results of an extensive 
survey of GC CMDs which may conform to the above definitions 
of HB bimodality and gaps. We believe that this compilation 
and the accompanying discussion will prove useful for future 
investigations of HB bimodality and gaps. 

We shall now proceed to describe in detail our theoretical 
efforts to model the HB of NGC 6229, beginning with a discussion
of the theoretical framework used in our analysis.  

\section{Theoretical Framework: Evolutionary Tracks and 
Synthetic HBs}

\subsection{Stellar Models and HB Simulations}

The evolutionary tracks for this project were computed with the same
numerical procedures as in Sweigart \& Gross (1976, 1978), except for
the use of updated input physics as described by Sweigart (1997a, 1997b). 
A solar-scaled metallicity of $Z = 0.001$ was adopted, as seems
appropriate for NGC 6229 (${\rm [Fe/H]} = -1.44$; Harris 1996), after
assuming a slight enhancement of the $\alpha$-elements (Salaris,
Chieffi, \& Straniero 1993). The standard value for the main-sequence
(MS) helium abundance, as derived for Galactic GCs from the $R$-method 
(e.g., Buzzoni et al. 1983), was assumed, i.e., $Y_{\rm MS} = 0.23$. 

An evolutionary track for a mass of $0.82\,M_{\sun}$ was first
evolved up the red giant branch
(RGB) without mass loss and then through the helium
flash to the zero-age HB (ZAHB).
This choice for the mass corresponds to an age at the tip of the
RGB of 15 Gyr. Lower mass ZAHB models were then generated by
removing mass from the envelope of this $0.82\,M_{\sun}$ ZAHB
model. The minimum ZAHB mass was $0.497\,M_{\sun}$, which only
slightly exceeds the core mass 
$M_{\rm c} = 0.4937\,M_{\sun}$ for the adopted
chemical composition. In total, a grid of 16 HB tracks was
computed. This grid was supplemented by two additional HB
tracks for $M = 0.495\,M_{\sun}$ and $0.496\,M_{\sun}$ for
producing Fig. 15 only. Some canonical HB
tracks for $Y_{\rm MS} = 0.23$, $Z = 0.0005$ from Sweigart (1997a,
1997b) were also used in preparing Fig. 3.

The above HB evolutionary tracks were fed into an updated version of
Catelan's (1993) HB synthesis code. We refer the reader to this 
paper for a description of the method. However, several improvements 
to this code deserve mention.

The prescriptions of Kurucz (1992) have been
adopted to transform from the theoretical
($\log\,L$, $\log\,T_{\rm eff}$) to the observational ($M_V$, $\bv$) 
plane. Silbermann \& Smith (1996) have discussed the possible need 
for a shift of the zero point of the Kurucz bolometric corrections (BCs). 
No such shift was applied in the present study, since it is unclear 
that there is a solid physical motivation for
applying the same ``zero-point" correction to stars of widely 
different temperatures, gravities and metallicities, and thus of 
widely different atmospheric structures (but see Buser \& Kurucz
1978, especially their Sec. V, for a more detailed discussion
of the ``arbitrariness" of the zero points of BC scales).
Hill's (1982) Hermite
interpolation algorithm has been employed for interpolation with
respect to $\log\,g$ and $\log\,T_{\rm eff}$. When necessary, 
linear interpolation was adopted with respect to [m/H].

Random numbers were generated using the procedure 
described by Wichmann \& Hill (1987). For the generation of 
Gaussian deviates, the Box-M\"uller method (see description in 
Bevington \& Robinson 1992) has been adopted. This method 
reliably describes the {\em tails} of Gaussian distributions,
which may be particularly important 
in the modelling of scarcely populated, extended blue HBs.
It is important to bear in mind the subtle, but 
fundamental, difference between a {\em Gaussian deviate} and a 
{\em Gaussian distribution}. Deviates are randomly-generated 
distributions which asymptotically approach the true
distribution it is meant to represent in the limit 
$N \rightarrow \infty$. 
In CMDs of Galactic GCs, the total number of observed HB stars is 
often $N_{\rm HB} \lesssim 200$ (see, e.g., Table 1 in Lee et al. 
1994). For samples of this size, one can expect the distinction
to become important (see, e.g., Newman, Haynes, \& Terzian 1989). 
In the remainder of this paper, however, the more familiar 
term ``Gaussian distribution" will be used interchangeably with 
``Gaussian deviate."

Observational errors in $B$ and $V$, assumed to be Gaussian, 
have also been introduced using the Box-M\"uller method.

The synthetic HB code has also been adapted to generate bimodal 
distributions in mass. Instead of being characterized by three 
free parameters, $\langle M_{\rm HB} \rangle$ (mean mass), $\sigma_M$  
(width of the Gaussian mass distribution) and $N_{\rm HB}$ (total 
number of HB stars), as in the unimodal case, the bimodal simulation 
requires {\em six} free parameters: two 
($\langle M_{\rm HB} \rangle$, $\sigma_M$) pairs, 
$N_{\rm HB}$, and a normalization factor $f_1$ that gives the
fraction of the $N_{\rm HB}$ stars contributed by the first Gaussian. 

The existence of a photometric limit in the CMD below 
which the incompleteness becomes $100 \%$ has been accounted for in the
following way. After generating a simulated CMD in the standard way, 
a search is performed for any stars fainter than $\Delta V^{\rm RR}_{\rm lim}$ 
below the mean RR Lyrae level. Such stars are then discarded, and 
``replacement" stars are generated from the same original mass 
distribution. The process is iterated until all $N_{\rm HB}$ stars lie
above the brightness limit. 

Finally, in the present computations a width 
$\Delta \log\,T_{\rm eff} = 0.075$ has been adopted for the instability 
strip. 

\subsection{The $\ell_{\rm HB}$ Coordinate}

A numerical routine has also been added to the HB synthesis code that
automatically computes the $\ell_{\rm HB}$ coordinate, which is linear 
along the HB ridgeline (Ferraro et al. 1992a, 1992b; Dixon et al. 1996).
Coordinates of this type are very useful for studying the presence of
gaps on the HB (Crocker et al. 1988; Rood \& Crocker 1989), since they 
effectively remove the degeneracy in HB colors for $(\bv)_0 \lesssim 0$
(cf. Fig. 16 below).

To compute the HB ridgeline, we make use of the synthetic HB code
to generate a ``fiducial" HB model with 8,000 stars. Mean colors and 
magnitudes are then computed at selected points along the HB, and 
a cubic-spline curve fitted to these points. (One ridgeline thus 
obtained can be seen in Fig. 8a below.)

After projecting each star onto the ridgeline, a length value 
is then computed along it, from the star's location up to 
the red limit of the distribution. The scaling relations
$\Delta\ell_{\rm HB}/\Delta (\bv)_0 \simeq 22.24$ and
$\Delta\ell_{\rm HB}/\Delta M_V \simeq 4.63$ are assumed,
which are similar to those obtained by Dixon et al. (1996) 
for M79 = NGC 1904. 

In Fig. 1, we show the HB region of the CMD of NGC 6229, taken from 
Paper I. Using the above procedure, we obtained the NGC 6229 
$\ell_{\rm HB}$ distribution from these data. A total of 31 RR Lyrae 
variables was assumed to be present---the maximum 
number suggested in Paper I---and their colors were 
assumed to be distributed randomly inside the instability strip 
boundaries given by Carney et al. (1991). The resulting 
$\ell_{\rm HB}$ distribution is shown in Fig. 2. Note that the 
maximum value attained by $\ell_{\rm HB}$ corresponds very 
closely to the $L_t$ value provided by Fusi Pecci et al. (1993)
for this cluster, $L_t = 14$.

The bimodal nature of the NGC 6229 CMD is also evident in this plot.
One clearly recognizes the red HB ``clump" in the
$0 \leq \ell_{\rm HB} \lesssim 3$ region. The blue tail corresponds 
to the well-populated region at $\ell_{\rm HB} \gtrsim 7$. 
In between, one finds a ``valley" in the distribution. 
The gap on the blue HB is 
easily recognized, lying at $\ell_{\rm HB} \simeq 8.8$, with width 
$\Delta\ell_{\rm HB} \approx 0.5$. We will analyze 
the statistical significance of this feature in Sec. 5 below.

\section{Synthetic HBs for NGC 6229}

It has sometimes been suggested that bimodal HBs may
be produced by unimodal mass distributions along the HB
(Norris 1981; Lee et al. 1988; Walker 1992). In the present
section, we will investigate this possibility for the case of
NGC 6229 on the basis of detailed HB simulations.

We begin by displaying some 
key HB morphology parameters for NGC 6229 in 
Table 1. The number counts have been corrected for 
completeness as described in Paper I. The following 
quantities are given: the Mironov (1972) parameter $B/(B+R)$; 
the Lee-Zinn parameter $(B-R)/(B+V+R)$ (cf. Lee et al. 1994); the 
Buonanno (1993) parameter $(B2-R)/(B+V+R)$; the number ratios 
$B: V: R$, obtained assuming that approximately half of the new
candidate RR Lyrae variables (Paper I) are ``real;" the total length 
of the HB $L_t$ (Fusi Pecci et al. 1993), computed as in 
Sec. 3.2 above; and, finally, the color of the red edge of the 
HB, $H\!B_{\rm RE}$ (cf. Fusi Pecci et al. 1993).
In the simulations, $H\!B_{\rm RE}$ was defined as the mean 
$(\bv)_0$ color for the reddest $2.5 \%$ of the HB stars. 

Our goal was to select synthetic HBs with unimodal and bimodal mass 
distributions (UMD and BMD, respectively) that would best match 
{\em all} of these HB morphology parameters. We have run several
hundred test cases with large ($\sim 500 - 5,\!000$) populations 
of HB stars to narrow down the allowed range of free parameters 
that permit a satisfactory global fit to the observed 
HB morphology. In the BMD case, the search 
was further constrained by the (arbitrary) requirement that 
$\sigma_{M,1} \approx \sigma_{M,2}$. We also assumed
that $\Delta V^{\rm RR}_{\rm lim} = 2.2$ mag (Paper I).

This search yielded the following two simulations:

\noindent
$\begin{array}{lll}
          {\rm UMD:} & \langle M_{\rm HB} \rangle = 0.5595\, M_{\sun}, &
                       \sigma_M = 0.1\, M_{\sun}; \\
          {\rm BMD:} & \langle M_{\rm HB,1} \rangle = 0.59\, M_{\sun}, &
                       \sigma_{M,1} = 0.025\, M_{\sun}, \\
                     & \langle M_{\rm HB,2} \rangle = 0.67\, M_{\sun}, &
                       \sigma_{M,2} = 0.025\, M_{\sun}, \\
                     & f_1 = 0.578.
 \end{array}$ 
            
\noindent For these two selected cases, we then
computed two series of synthetic HBs, each containing
1,000 simulations with $N_{\rm HB} = 235$ (Paper I) stars. 
Observational scatter was also added, assuming Gaussian 
error distributions with $\sigma_V = 0.017$ mag and 
$\sigma_B = 0.031$ mag, as suggested by the data in Paper I. 
Variable and non-variable HB stars were treated in exactly the 
same way, as far as the observational errors are concerned; we
thus expect these HB simulations to show significant overlap in color 
between the variable and non-variable stars.
The average values of the HB morphology parameters,
as well as the corresponding standard deviations,
are given in Table 1 for both the UMD and BMD cases.

Inspection of Table 1 discloses that {\em both the UMD and
BMD cases are quite successful at reproducing, to within the 
errors, all of the major HB morphology parameters for NGC 6229}.
In the UMD case, however, this has required a 
large value for the HB mass dispersion. 
Our test runs show that significantly smaller $\sigma_M$ 
values (but still larger than the typical values listed, e.g., 
by Catelan 1993) would still reproduce the observed
HB morphology parameters, except
for $(B2-R)/(B+V+R)$, which provides a stringent constraint on the
allowed simulations by demanding that a large fraction of the blue 
HB population lie at $(\bv)_0 < 0.02$ mag. 
These results indicate that {\em HB bimodality is consistent 
with a unimodal distribution in mass, provided the mass dispersion
is sufficiently large} (for further discussion see Sec. 7 below). 

The size of the mass dispersion required for bimodality
is, however, a function of the adopted metallicity. This 
is shown in Fig. 3, where the run of $V/(B+V+R)$ as a function of 
$\sigma_M$ is compared with those of $B/(B+V+R)$ and $R/(B+V+R)$ 
for two metallicities. In this plot, the Lee-Zinn 
HB morphology parameter was held fixed at the value appropriate for 
NGC 6229. The displayed quantities were obtained by averaging
over $\sim 10$ simulations with $N_{\rm HB} \simeq 1,\!000$ for
each $\sigma_M$ value. A larger $\sigma_M$ value is clearly 
required in the lower $Z$ case to produce HB bimodality (as defined 
in Sec. 2). This behavior can be 
straightforwardly understood if we recall that the HB effective
temperature becomes less sensitive to changes in mass as the 
metallicity decreases [see Fig. 7 in Buonanno et al. (1985)].

Fig. 4 shows twelve cases selected at random from the series of 1,000 UMD 
simulations. The reason for displaying several of these simulations 
in one single plot is to highlight the effects of the statistical 
fluctuations upon the HB morphology. We emphasize that all 
displayed models are based on exactly the same underlying physical 
assumptions and input parameters. A similar random sample of 12 models for 
the BMD case is displayed in Fig. 5.

Histograms showing the $\ell_{\rm HB}$ and $\log\,T_{\rm eff}$ 
distributions for the UMD simulation in Fig. 4b
are given in Fig. 6. Fig. 7 shows similar histograms for 
the BMD simulation in Fig. 5f. (These two examples were chosen randomly 
from the set of simulations presented in 
Figs. 4 and 5.) From Fig. 6, one sees that {\em a bimodality 
in $\ell_{\rm HB}$, or even in $\log\,T_{\rm eff}$, does not 
necessarily imply a true bimodality in the underlying mass 
distribution}. For a sufficiently wide UMD, one peak in the 
temperature distribution will naturally occur at the red end of the HB,
where the evolutionary tracks for the higher masses
tend to pile up. Unless it is erased by
statistical fluctuations, a peak on the blue HB will also 
be naturally present, {\em provided} the peak of the UMD 
corresponds to a ZAHB point on the blue HB. 

A quick inspection of Figs. 4 and 5 suggests that it may be
difficult to discriminate between the UMD and BMD cases. A closer
analysis reveals, however, that the blue HB in the 
UMD case tends to be more uniformly populated than in the 
BMD case due to the very large $\sigma_M$ value 
required in the UMD case. This effect is indeed suggested in 
Figs. 6 and 7. In fact, a two-sample Kolmogorov-Smirnov test 
applied to the 1,000 UMD and 1,000 BMD simulations
slightly favors the UMD case, in the mean, as the 
actual representation of the observed $\ell_{\rm HB}$ distribution in 
NGC 6229 (Fig. 2). However, the difference is not very
large, and we do not consider it significant for the following
reasons: i) The synthetic HBs contain more blue HB stars than does
the CMD in Paper I, because the completeness corrections discussed 
in Paper I have been taken into account in the {\em number counts} 
only, and ii) The precise number and color distribution of the RR Lyrae 
stars in the observed CMD are uncertain.

The remarkable similarity between the UMD and BMD simulations
can be largely ascribed to the adopted brightness limit,
$\Delta V^{\rm RR}_{\rm lim} = 2.2$ mag. 
In fact, the very large mass dispersion in the UMD case leads to  
a {\em very} extended blue tail. This point is illustrated in Fig. 8 by
two sample simulations with $N_{\rm HB} = 300$, computed for the UMD case with 
$\Delta V^{\rm RR}_{\rm lim} \rightarrow \infty$. 
Mean HB morphology parameters over 12 such simulations 
are given in Table 2. 

As one can see, the distribution may reach as hot
as the extreme HB (EHB) region. The $\ell_{\rm HB}$ and $\log\,T_{\rm eff}$ 
distributions for the simulation in Fig. 8b, plotted in Fig. 9, 
shows a significant number of EHB stars. However, it is important to note
that a lower value for $(B2-R)/(B+V+R)$ would produce a less extended
blue HB tail. In the BMD case, 
on the other hand, a very extended blue tail is not {\em necessarily} 
implied, as shown in Fig. 10, although a long blue tail might 
also be present in this case, if the mass dispersion
for the low-mass mode is larger than assumed here. In fact, as 
discussed in Paper I, there are some indirect indications that a 
well-populated blue tail may be present in NGC 6229. 

Finally, it is important to note that the horizontal part of the
synthetic HBs seems to be quite ``flat." The observations,
on the other hand, indicate that the red 
end of the blue HB may be brighter than the main portion 
of the red HB (Paper I), as previously suggested by Stetson 
et al. (1996) for NGC 1851 and NGC 2808. If real, this would 
imply that, in all likelihood,
a different parameter besides the mass 
varies between the blue and the red HB. One natural 
candidate is the envelope helium abundance (Sweigart 1997b). 
{\em We emphasize the importance of constraints of this type for 
assessing the cause of HB bimodality on a cluster-by-cluster basis}.

\section{On the Reality and Nature of HB Gaps}
A particularly important aspect of
Figs. 4 and 5 is the surprisingly large fraction of the HB
simulations that display prominent gaps. In this section, we will
present a new method for analyzing the statistical significance
of such gaps. We begin with the standard analysis of the gap found 
on the {\em observed} $\ell_{\rm HB}$ distribution of NGC 6229. 
This gap is clearly
present in Fig. 2 at $\ell_{\rm HB} \simeq 8.8$, and has a size
$\Delta\ell_{\rm HB} \simeq 0.5$. 

\subsection{The Gap on the Observed HB}

The standard method for evaluating the statistical significance of
gaps on the HB is the one advanced by Hawarden (1971) and used by
many other researchers (e.g., Newell 1973; Crocker et al. 1988).
The method works in the following way: from the average slope 
of the cumulative distribution of stars to the blue and to the
red of a detected gap, and from its estimated width, one computes
the ``expected" number of stars inside the gap, $N_{\star}$. 
From the number of stars which are {\em actually}
found inside the gap, $N_{\rm gap}$ ($= N_0$ in Hawarden's 
notation), the following ``$\chi^2$ statistic"
is then formed:

\begin{equation}
 \chi^2 = \frac{(N_{\star} - N_{\rm gap})^2}{N_{\star}}.
\end{equation}

\noindent Standard statistical tables (e.g., Abramowitz \& Stegun 
1965, Table 26.7) are then used to evaluate the
``statistical significance" of the detected feature. Although
not explicitly stated, apparently the assumed number of degrees
of freedom in Hawarden's test is $\nu = 1$, for an example 
is given of a case where $\chi^2 = 8.5$, whose probability of 
being ``an accidental result is about 0.5 per cent." That 
applications of the method generally assume $\nu = 1$ 
is also supported by the values provided by Newell in 
his Table 1, and by Crocker et al. in their Table 1, although
we have been unable to reproduce the probability value in the
latter study for NGC 1851.

If we apply the same method to NGC 6229, we find that
$\approx 14$ HB stars are expected inside the gap region,
whereas only $\approx 3$ are found (cf. Figs. 1 and 2). Thus,
$\chi^2 = 8.64$, which, for $\nu = 1$, would imply a $99.7 \%$
probability that the gap is not a statistical fluctuation. 
How does this compare with the results of the Monte Carlo 
simulations?

\subsection{Gaps on the Synthetic HBs}
If the probability that the NGC 6229 gap is not a statistical
fluctuation is indeed as high as $99.7 \%$, one would expect only 
about 3 out of every 1,000 simulations to present a ``NGC 6229-like" 
gap on the blue HB. To test this prediction, 
we have carefully inspected each of our 1,000 simulations in
the ($M_V$, $\bv$) plane
for the presence of a gap on the blue HB. We find that 
$\sim 6 \% - 9 \%$ of all simulations show gaps which are
comparable in size and/or sharpness with the observed one. 
{\em This stands in sharp 
contrast with the negligibly small fraction 
that would be expected on the basis of Hawarden's 
(1971) method}. In fact, if gaps which are a little less 
sharp or narrower are also included, we find that up to 
$\approx 15 \%$ of all simulations show gaps on the blue HB. 
These numbers increase further by a few percent if  gaps 
inside the instability strip or on the red HB are also included.

Selected synthetic HBs with quite prominent gaps 
are shown in Fig. 11 (UMD case) and in Fig. 12 (BMD case).
These were chosen from the same two sets of 1,000 simulations
previously discussed. As one can see, prominent gaps can 
indeed appear virtually anywhere along the HB.

These Monte Carlo simulations suggest that 
Hawarden's (1971) method substantially overestimates the
probability that gaps are ``real." To understand the reason
for this, we recall that Hawarden's method attempts to
answer the following question: What is the probability that
a gap {\em at the observed location} is real? By requiring
a gap to be at the observed location, this method effectively
constrains the probability that a gap can arise from a 
statistical fluctuation. There is, of course, no way of
knowing a priori where a statistical fluctuation might
occur. Consequently the more fundamental question to ask is:
What is the probability that a gap will be found {\em somewhere}
along the HB? This latter question provides an additional
degree of freedom to the statistical analysis and therefore
a more realistic estimate of the importance of statistical
fluctuations.

The present situation for the HB gaps resembles the 
controversy that arose some time ago regarding the 
significance of gaps on the RGB. Sandage, 
Katem, \& Kristian (1968) first reported that a
``major significant" gap was present on the RGB of M15,
and subsequently similar features were
detected in several other GCs. However, 
Monte Carlo experiments by Bahcall \& Yahil (1972) showed 
that RGB gaps occurred far more frequently than expected 
from the Sandage et al. statistical 
arguments. As described by Renzini \& Fusi Pecci 
(1988) some twenty years later, ``the tendency (\ldots) has 
been for such [RGB] gaps to get filled with increasing sample 
size, as simulations had anticipated (Bahcall \& Yahil 1972)." 
Newman et al. (1989) discuss a similar problem 
in the study of the periodicities in galaxy redshifts. 

\subsection{On the Probability of Finding Gaps {\em Somewhere}
            along the HB}
We consider now a simple mathematical model, based on the
binomial distribution, for estimating the probability of
finding a gap somewhere along the HB.

Consider a hypothetical GC whose $\ell_{\rm HB}$ 
distribution is perfectly uniform. Let us
divide the HB into $N_{\rm b}$ 
bins of identical width $\Delta\ell_{\rm HB}$. 
We then place, at random, $N_{\rm HB}$ stars onto the HB 
of this cluster, and construct an $\ell_{\rm HB}$ distribution
for the stars in this sample. Clearly, the probability that
a {\em given} star will be placed in the $j$-th bin is 
$p_j = 1/N_{\rm b}$, since the distribution is known {\em a
priori} to be uniform. We ask the question: what is the probability 
$P_j$ that out of this sample of $N_{\rm HB}$ stars 
the total number of stars $N_j$ in the $j$-th bin 
will be less than or equal to some value $N_{\rm gap}$? The answer 
to this question is given by the binomial distribution:

\begin{equation}
 P_j = \sum_{k=0}^{N_{\rm gap}} 
       \left( \begin{array}{c}
                N_{\rm HB} \\ k
            \end{array} \right) 
        p_j^k (1-p_j)^{(N_{\rm HB} - k)},
\end{equation}

\noindent where $p_j = p = 1/N_{\rm b}$.

By applying this simple model to the blue HB of NGC 6229---which is equivalent 
to assuming, as a first crude approximation, that its ``true" underlying
$\ell_{\rm HB}$ distribution is uniform---we can
estimate the probability that a gap of the size actually present in
NGC 6229 will be seen in the $j$-th bin. (In fact, we have noted
before that the distribution of stars along the blue tail, for
a large mass dispersion $\sigma_M$, will indeed be quite uniform
in the mean.) For NGC 6229, the number of stars on the blue HB 
(corrected for completeness) is $N_{\rm BHB} = 147$ (Paper I).
Further, the blue HB distribution extends from
$\ell_{\rm HB} \approx 7$ to $\ell_{\rm HB} \approx 14$. 
As we have seen, the size of the observed NGC 6229 gap is 
$\Delta\ell_{\rm HB} \approx 0.5$, and the number of stars inside the 
gap is 3. If NGC 6229's blue-HB $\ell_{\rm HB}$ distribution is divided
into 14 bins of size $\Delta\ell_{\rm HB} = 0.5$, then
the probability that a given star 
will be found in the $j$-th bin is 1/14. We then find from 
equation (2) that the probability that the $j$-th bin will
contain 3 or less stars is

\begin{eqnarray}
 P_j & = & \sum_{k=0}^{3}
           \left( \begin{array}{c}
                    147 \\ k
                  \end{array} \right)
           \left(
                    {{1} \over {14}}
           \right)^k
           \left(
                    1 - {{1} \over {14}}
           \right)^{(147 - k)}  \nonumber   \\
     & = & 0.58 \%.
\end{eqnarray}

\noindent The probability of finding a gap as sharp as, or sharper than, the
one found in NGC 6229 {\em in the $j$-th bin} is indeed quite low. 
However, as already anticipated, 
this calculation does not give an answer to 
the {\em really} important question, namely: What is the probability 
$P_{\rm gap}$ that a sharp gap will be found {\em somewhere} 
along the blue HB?

We can apply the above formalism to answer this question. The total
number of bins is 14, and, from equation (3), the probability of a gap 
occurring in the $j$-th bin is $P_j = 0.0058$. Thus from the binomial 
distribution the probability that a gap will be found 
{\em somewhere} along the blue HB is given by

\begin{eqnarray}
 P_{\rm gap} & = & 
           \left( \begin{array}{c}
                    14 \\ 1
                  \end{array} \right) 
           0.0058^1 \, (1 - 0.0058)^{(14 - 1)} \nonumber \\ 
     & = & 7.5 \%.
\end{eqnarray}

\noindent Thus, the probability that a gap will be present
on the blue HB is {\em at least an order of magnitude 
larger than commonly realized} (Hawarden 1971; Newell 1973; Crocker
et al. 1988). Although this rough estimate is based on a crude model 
that can certainly be refined with a more realistic approximation for 
the probability function $p_j$, it is in reasonable agreement with 
the results of our detailed Monte Carlo experiments.

Does this imply that all the known HB gaps (cf. Sec. 2) are caused
by statistical fluctuations? While the probability that a 
blue HB gap is due to a statistical fluctuation is at least
an order of magnitude larger than previously 
realized, this does not imply that some, and perhaps all,
of the detected gaps are not real in some deeper physical sense.  
This issue can only be settled by a very careful case-by-case 
study of the available observational data. For instance, the
occurrence of gaps at very similar positions on the HB in
clusters of nearly identical metallicity might be suggestive
of a genuine physical mechanism for producing gaps.
Therefore, besides reassessing the statistical significance 
of each individual gap, it would be very important to apply 
some of the proposed tests which are capable of constraining 
their possible physical causes (Rood \& Crocker 1985a, 1985b; 
Crocker et al. 1988; D'Cruz et al. 1996; Sweigart 1997b). 
We intend to employ both canonical and
non-canonical stellar evolution models, coupled with Monte Carlo 
techniques, in order to attempt to make some progress in this 
direction (Catelan \& Sweigart 1997).

\section{Synthetic HBs for NGC 1851 and NGC 2808}
We have shown that the UMD simulations can satisfactorily 
account for the HB bimodality found in NGC 6229, provided
the mass distribution is substantially wider than commonly
assumed for the GCs. As a further step in this
investigation, we now consider whether a similarly wide
mass distribution might be responsible for the HB bimodality
in the archetypal clusters NGC 1851 and NGC 2808.

\subsection{On the HB Bimodality in NGC 1851}
Table 3 gives the HB morphology parameters for NGC 1851. These
parameters are based on the following number counts derived
from Fig. 7 in Walker (1992): $B = 49$, $V = 16$, $R = 96$. 
We have also used $B2/(B+V+R) = 0.17$ from Buonanno et al. (1997)
and the $L_t$ and $H\!B_{\rm RE}$ values from Fusi Pecci et al. (1993).

We have performed a search for a UMD simulation that would optimally 
match these HB morphology parameters. A simulation characterized by
$\langle M_{\rm HB} \rangle = 0.665 \, M_{\sun}$,
$\sigma_M = 0.055 \, M_{\sun}$ was 
found to give a satisfactory fit. We then computed 
a sequence of 100 simulations with $N_{\rm HB} = 161$, assuming an 
observational scatter of  $\sigma_{B,V} < 0.01$ mag.
The resulting mean values for the HB morphology parameters 
are also given in Table 3. Two representative simulations are
shown in Fig. 13. As one can see, the match to the observed HB
parameters is quite satisfactory, and we conclude that the HB bimodality 
of NGC 1851 may also be caused by a unimodal, if peculiarly wide, HB mass 
distribution.
Walker (1992) has noted that the red HB is more sharply peaked
in color than would be predicted by simulations similar to
those in Fig. 13. However, the HST CMD for NGC 1851 by
Sosin et al. (1997b) seems to show a red HB that resembles 
more closely the morphology expected from the present simulations.

The observed and expected 
$L_t$ values in Table 3 seem to differ by a 
substantial amount---although we cannot guarantee that the
two quantities are on exactly the same scale, even though
our calibration reproduces the Fusi Pecci et al. (1993)
$L_t$ value for NGC 6229 (cf. Sec. 4).
The reason for this disagreement can be understood by comparing
the two simulations in
Fig. 13. In panel a, a simulation with a relatively ``short" 
blue tail is displayed; this is similar to the CMD presented by
Walker (1992) in his Fig. 6, as well as the one previously
published by Stetson (1981)---the latter having been used
by Fusi Pecci et al. in their measurement of $L_t$. In about 
2 out of every 3 UMD simulations, however, we find
a dribble of quite hot stars, lying at $M_V > 3$ mag, 
which increase the predicted value of $L_t$.
A few such hot stars are indeed suggested by the CMDs
in Walker's Fig. 5 (see also Saviane et al. 1997). The 
{\em Ultraviolet Imaging 
Telescope} (Parise et al. 1994; Landsman 1994) did detect two very 
hot HB stars in NGC 1851, one of which Landsman has confirmed to 
be a radial velocity cluster member. Additional hot HB stars
may be present closer to the center of this very concentrated 
cluster, where crowding prevented Parise et al. from obtaining 
reliable UV colors. The present simulations suggest that these 
stars may be compatible with a standard, if quite wide, UMD.

\subsection{On the HB Bimodality in NGC 2808}
We have also computed some simulations to see if the 
HB bimodality in NGC 2808 could also be reproduced with a 
UMD. We employed the number counts from Sosin et al. (1997a)
to estimate the HB morphology parameters for this cluster.
Some 800 HB stars were measured in NGC 2808 by these authors 
using the HST, yielding one of the most spectacular CMDs for 
a GC ever obtained. The number of RR Lyrae variables in 
NGC 2808 seems to be very low (Clement \& Hazen 1989), 
although claims have on occasion been raised that the 
instability strip could be more substantially populated 
(Byun \& Lee 1991). In the HST field, however, the number
of RR Lyrae variables seems to be safely constrained to
be $< 20$, and possibly much lower (Dorman
1997). This gives  $V/(B+V+R) = 0.024$ as an upper limit 
to the fraction of RR Lyrae variables in NGC 2808.

We were totally unable to obtain a satisfactory UMD simulation
for NGC 2808. The minimum number of RR Lyrae
variables in our simulations occurs when the peak of the mass
distribution is placed at 
$\langle M_{\rm HB} \rangle \approx 0.497 \, M_{\sun}$,
with an exceedingly large
mass dispersion $\sigma_M > 0.3\, M_{\sun}$; even
in this (probably highly artificial) case, however, the 
RR Lyrae fraction is as high as $V/(B+V+R) \simeq 0.06$. 
In fact, in a series of 100 synthetic HBs for these
UMD parameters (where observational scatter $\sigma_{B,V} < 0.015$ 
mag was also added), the minimum RR Lyrae fraction was
$V/(B+V+R) = 0.039$. We show the simulation with the minimum 
number of variables in Fig. 14a, and another simulation with 
a curious gap on the blue HB in Fig. 14b. The agreement between 
these simulations and the CMD presented by Sosin et al. (1997a) 
is not satisfactory. It appears therefore that a bimodal, or
possibly multimodal, mass distribution is required to fit the
NGC 2808 HB.

\section{Discussion}
With the insight provided by our HB simulations, we shall now
analyze some previously proposed scenarios for the origin of HB 
bimodality and gaps. For a discussion 
of non-standard scenarios, we refer the reader to the recent
papers by van den Bergh (1996), Catelan (1997), Sosin et al. 
(1997a), and Sweigart (1997b).

\subsection{HB Bimodality}
It has sometimes been suggested that bimodal HBs may
be accounted for by unimodal mass distributions along the HB
(Norris 1981; Lee et al. 1988; Walker 1992). In particular,
Walker has argued that, given a sufficiently wide dispersion
in mass along the HB, bimodality is naturally implied. But
does this apply equally well to both $\bv$ colors and
temperatures?

In Figs. 15a and 15b we show the ($\log\,L$, $\log\,T_{\rm eff}$) 
and ($M_V$, $\bv$) planes, respectively, for a ``reference" 
synthetic HB populated by a large number (100,000) 
of HB stars randomly sampled from a uniform distribution 
in mass along the ZAHB over the range $0.495 \leq M/M_{\sun} \leq 0.820$.
This large number of HB stars was used to minimize any statistical 
fluctuations. Fig. 15b shows that, on the observational diagram, the 
number of stars at intermediate $\bv$ colors (where the RR Lyrae 
variables are found) seems to be significantly depleted in comparison 
with the number of stars both on the blue HB and on the red HB. 
In Fig. 16, we show the color-temperature relation from
the Kurucz (1992) model atmospheres for a representative gravity
and metallicity. These diagrams show very clearly the 
well-known fact (e.g., Rood \& Crocker 1989) that bimodal 
color distributions may be produced due to the piling up 
of HB evolutionary tracks at the low-temperature (i.e., 
high-mass) end, and to the insensitivity of the $\bv$ color
to temperature changes for $\log\,T_{\rm eff} \gtrsim 3.9$. Thus,
HB bimodality---at least in the observational sense previously
defined---is naturally implied for sufficiently wide mass 
distributions, as we had indeed found on the basis of detailed 
simulations. However, inspection of Fig. 15a reveals that, 
for a uniform mass distribution, a rather uniform distribution 
in temperatures results. 

Interestingly, Walker (1992) has
proposed that the morphology of the HB evolutionary tracks---in
particular, the presence of long ``blue loops" in the post-ZAHB 
evolution---might produce a bimodal distribution in 
temperature (and thus also in the number counts and $\bv$ colors) 
along the HB. The only requirement, according to Walker, 
is that the mass dispersion be substantially larger than
typically assumed in HB simulations. Can track morphology 
really lead to two main statistical modes in the HB temperature 
distribution? To address this question, we show in Fig. 17 a 
synthetic HB employing 20,000 stars randomly sampled from a
uniform mass distribution over a mass range
chosen to cover the temperature range in Walker's 
Fig. 8 for NGC 1851. Our assumption of a flat mass distribution
favors Walker's suggestion by adding more blue HB stars than
in the detailed simulations for this cluster shown in Fig. 13.
The same chemical composition as in Walker's study was also adopted. 
Two histograms for $T_{\rm eff}$ are given in Fig. 17: one which
includes the canonical evolution of the HB stars away from the
ZAHB, and another where all evolution away from the ZAHB is
suppressed. As one can clearly see, both histograms are
comprised of a single statistical mode on the red clump 
region, accompanied by a quite uniformly populated 
distribution extending to higher temperatures. There 
is no sign of a peak in the temperature distribution 
blueward of the instability strip. These histograms clearly show
that evolution off the ZAHB does not help in producing two
main modes in the temperature distribution.

We have carried out a similar simulation using the HB 
evolutionary tracks of Lee \& Demarque (1990), as shown in
Fig. 18. As remarked
by Catelan \& de Freitas Pacheco (1993), the Lee \& Demarque 
HB evolutionary tracks, for some still unclear reason, 
present blueward loops which are much longer
in $\Delta\log\,T_{\rm eff}$ (for the same chemical composition)
than in other independent HB evolutionary calculations.
This difference in track morphology dramatically affects the
synthetic HBs, particularly the luminosity width, as can clearly
be seen by comparing Figs. 17 and 18.
Interestingly, the more recent Koopmann et al. (1994) and 
Yi, Lee, \& Demarque (1995) HB tracks show blueward loops 
which are in much better agreement with those presented by, 
e.g., Sweigart (1987), as well as with those employed in the
present work. The differences in track morphology 
notwithstanding, this simulation also shows a remarkably flat
temperature distribution blueward of the instability strip.
(The slight irregularities in this HB simulation are due to
the coarseness of the Lee \& Demarque grid.) 
The histograms once again demonstrate that the HB 
evolution does not contribute to the production of two main
statistical modes in the temperature distribution.
Our results do not therefore support the argument (Lee et al. 1988) 
that bimodality can arise naturally from ``evolution away 
from the ZAHB."

There is, however, one circumstance under which the blueward loops 
may be {\em directly} responsible for producing HB bimodality. For 
some combinations of evolutionary parameters---for instance, for 
large values of the envelope helium abundance, or for low values of
the helium-core mass $M_{\rm c}$---blueward loops may become 
{\em very} long indeed (see, e.g., Sweigart \& Gross 1976). Primordial 
(Shi 1995) or evolutionary (Sweigart 1997a, 1997b) processes might be
responsible for a large helium abundance in HB stars. A substantial 
decrease in $M_{\rm c}$ would, however, be harder to justify
(Sweigart 1994; Catelan, de Freitas Pacheco, \& Horvath 1996). 
In either case, for sufficiently large $Y$ or low $M_{\rm c}$ 
values ZAHB stars on the red HB clump
will evolve along extremely long blueward loops and thereby produce 
a substantial number of blue HB stars, and possibly even 
a relatively long blue tail.
Moreover, provided the ZAHB mass distribution
does not populate the region between the red clump and the blue HB, 
few stars would be present near the instability strip, 
because of the rapid evolutionary 
pace at these intermediate temperatures. If such a
scenario were to be responsible for a bimodal HB, however,
we would expect the HB distribution to be {\em sloped} on the 
CMD (Catelan \& de Freitas Pacheco 1996). Interestingly, 
bimodal {\em and sloped} HBs have recently been reported for 
the metal-rich GCs NGC 6388 and NGC 6441 (Piotto et al. 1997;
Rich et al. 1997). We have carried out extensive simulations to
explore this scenario in detail and have confimed that high
$Y$ values can lead to both bimodal and sloped HBs. Preliminary
results for these simulations have been presented by 
Sweigart \& Catelan (1997a) and will be discussed more fully
in a future paper (Sweigart \& Catelan 1997b).

\subsection{Gaps on the Blue HB}
The change in the morphology of canonical post-ZAHB evolutionary 
tracks as a function of mass has sometimes been invoked to explain
the presence of blue-HB gaps (e.g., Norris et al. 1981; Lee et 
al. 1994). In particular, it has been argued that the NGC 288 gap, 
and possibly the NGC 6752 one, may be easily explained in this way
(Lee et al. 1994). However, numerical experiments
by Castellani et al. (1995) failed to provide conclusive 
evidence for or against this suggestion.

Inspection of Figs. 15a and 15b shows no significantly 
underpopulated regions on the blue HB in either 
diagram. Similar plots using the 
Lee \& Demarque (1990) HB tracks do not show any 
significantly underpopulated regions either. This indicates 
that, contrary to previous claims, {\em gaps on the blue 
HB are not caused by the morphology of canonical HB tracks}.

One noteworthy feature in Fig. 15b is the non-monotonic behavior of
the number counts as a function of \bv at $\bv \approx 0.2 - 0.3$.
No corresponding feature seems to be present in the theoretical
plane (cf. Figs. 15a and 17). Inspection of the Kurucz (1992) color 
transformations (cf. Fig. 16)
discloses that this anomaly is related to an abrupt change in slope 
of the $(\bv ) - \log\,T_{\rm eff}$ relation at this region.
We refer the reader to the papers by Gratton, Carretta, \&
Castelli (1996, esp. their Sec. 5) and Lejeune, Cuisinier, \& 
Buser (1997) for discussions of this feature, which is not present 
in their more recent color-temperature transformations.

Some experiments with the
non-standard evolutionary tracks of Sweigart (1997a, 1997b) show 
that track morphology may play a r\^ole in generating gaps, 
but only if some physical parameter(s) (such as $Y$
or $M_{\rm c}$) changes {\em abruptly} as a function 
of temperature along the ZAHB, thus effectively generating a {\em break}
in track morphology at a given temperature, as opposed to the continuous
and smooth change that is found in the canonical case (Fig. 15).

A gap on the extreme HB is also predicted by the scenario outlined 
by D'Cruz et al. (1996). In their framework, very hot extreme 
HB stars ($\log\,T_{\rm eff} \approx 4.5$;
cf. their Fig. 2) are produced by the stars which undergo the He 
flash on the white dwarf cooling curve as a consequence of 
high mass loss rates on the RGB (Castellani \& Castellani 1993). 
In this case, however, the predicted gap location would be even
hotter than the gap in NGC 6752 (Sweigart 1997b).

\section{Summary}
The following are the main results of the present investigation:
\begin{itemize}
 \item HB bimodality may be caused by a unimodal distribution in mass,
       provided the mass dispersion on the HB is substantially larger
       than in ``typical" clusters. In particular, the HB bimodality 
       in NGC 6229 and NGC 1851 is
       quite satisfactorily accounted for by unimodal mass 
       distributions with mass dispersions 
       $\sigma_M = 0.1 \, M_{\sun}$ and $0.055 \, M_{\sun}$,
       respectively. In the NGC 6229 case, however, a unimodal
       mass distribution would imply the presence of
       a very extended and well-populated blue tail. 
       Further observations are urgently needed to check whether 
       such an extended HB is present or not. Our results do not
       support the claim (Lee et al. 1988) that HB bimodality
       arises from ``evolution off the ZAHB;"
 \item The NGC 2808 HB distribution cannot be reproduced by a unimodal
       mass distribution, but seems most decidedly to require a bimodal,
       or possibly multimodal, mass distribution;
 \item Extensive Monte Carlo simulations show that gaps along the HB
       are much more likely to occur as a consequence of statistical
       fluctuations than has been previously realized from the
       standard Hawarden (1971) and Newell (1973)
       $\chi^2$ technique. In particular, $\simeq 6\% - 9\%$ of
       our simulations show blue HB gaps resembling the one 
       observed in NGC 6229. In contrast, the 
       Hawarden-Newell technique predicts a probability of 
       $99.7 \%$ that the observed gap in NGC 6229 is real.
       The reason for this discrepancy can be traced to the fact
       that the Hawarden-Newell technique requires the statistical
       fluctuation to produce a gap {\em at the observed location}.
       As a result, the Hawarden-Newell method substantially
       overestimates the probability that a gap is real. The more
       fundamental question is whether or not statistical fluctuations
       can produce gaps of the observed size and/or sharpness 
       {\em somewhere} along the HB. The present investigation,
       in analogy with a previous study by Bahcall \& Yahill (1972)
       on the significance of gaps on the RGB, thus indicates that
       a comprehensive reassessment of the significance and nature
       of gaps on the blue HB is urgently required. In this regard,
       our study already shows that, contrary to previous claims 
       (e.g., Lee et al. 1994), gaps
       on the blue HB are {\em not} caused by the change in shape
       of the evolutionary tracks as a function of mass;
 \item We present a new mathematical model, based on the binomial
       distribution, for estimating the statistical significance
       of HB gaps and demonstrate that it predicts a probability
       for the reality of a gap that is consistent with our simulations;
 \item Refinement of the several possible {\em observational tests} 
       for the cause of
       HB bimodality and the presence of gaps along a 
       distribution (e.g., Crocker et al. 1988; Sweigart 1997b), 
       and their careful application to studies of each 
       individual cluster in which an anomaly of this sort has
       been reported, is thus {\em strongly encouraged}. 
       Upon this depends the reliable establishment 
       of the nature of the ``multiple parameters" 
       that seem to be affecting, to a 
       larger or smaller extent, the observed HB distributions
       in Galactic GCs (e.g., Buonanno et al. 1985;
       Rood \& Crocker 1985a, 1985b; Crocker et al. 1988; 
       Rood et al. 1993; Stetson et al. 1996; Sosin et al. 1997a; 
       Fusi Pecci \& Bellazzini 1997; Ferraro et al. 1997b;
       Sweigart 1997b; etc.).
\end{itemize}

\acknowledgments
M. C. would like to thank W. V. D. Dixon, B. Dorman and 
W. B. Landsman for useful information and discussions. 
This work was performed while M. C. held a 
National Research Council--NASA/GSFC Research Associateship. 
A. V. S. acknowledges support through NASA grant NAG5-3028.
This research was supported in part by the Bulgarian National 
Science Foundation grant under contract No. F-604/1996 with
the Bulgarian Ministry of Education and Sciences. 

\clearpage

\appendix

\centerline{{\bf Appendix}}

\vspace{1.0cm}

\centerline{{\bf A Compilation of (Candidate) ``Bimodal" and ``Gap" GCs}}

In this Appendix, we present a comprehensive compilation of GCs whose
HBs may satisfy the definitions of HB bimodality and gaps laid
out in Sect. 2. We hope this will provide a useful guide for future 
investigations of the problem of HB bimodality and gaps.

A list of clusters that appear to satisfy the definition of 
HB bimodality presented in Sect. 2 can be found in Table 4. 
In this table, values for the cluster metallicity 
${\rm [Fe/H]} = \log\,({\rm Fe/H})_{\rm GC} - \log\,({\rm Fe/H})_{\sun}$
and core concentration $c$ from Harris (1996; values adopted come from
the May 1997 issue of his catalogue) are given in the third and 
fourth columns, respectively. We list the $c$ values for these clusters
because of the suggestion that core concentration may be correlated with
anomalies in the HB morphology (e.g., Fusi Pecci et al. 1993; Castellani
1994 and references therein). The last column gives the
references where attention was drawn to the possible HB bimodality.

With respect to this table, the following additional
comments are in order. For NGC 6362, the suggestion of an HB
bimodality (Paper I)
was based on the Lee, Demarque, \& Zinn (1994) number counts, 
which in turn relied on Alcaino's (1972) 
color-magnitude diagram (CMD). This cluster has been more 
recently observed by Alcaino \& Liller (1986), but their
CMDs include only a few stars on the HB, so that a definite
answer regarding its HB bimodality must await a more extensive
investigation. For NGC 6712, the CMD published by Martins \& Harvel
(1981) may give some support for HB
bimodality, although the uncertainties remain large. The Lee et al.
number counts for this cluster, based on Sandage \& Smith's
(1966) CMD, give $B: V: R = 0.15: 0.11: 0.74$. New photometry for
NGC 6723 has very recently been presented by Fullton \& Carney (1996);
their CMD may give some support to the classification of this object
as a bimodal-HB cluster. For M75 = NGC 6864, new CCD investigations are 
especially encouraged, since the latest CMD available (Harris 1975) 
is strongly suggestive of an NGC 1851-like HB bimodality. Caputo
et al. (1973) have also suggested that M72 = NGC 6981 may have a
bimodal HB. However, the recent study by Kadla et al. (1995) has 
identified 9 possible new variable stars in the cluster. If
confirmed, this would rule out M72 as a bimodal-HB cluster, since
the number ratios would become
$B: V: R \simeq 0.33: 0.34: 0.33$ (Kadla et al. 1995). Note that we 
do not classify M5 = NGC 5904 as a bimodal-HB cluster, as opposed 
to Smith \& Norris (1983) and Lee et al. (1988), since 
the number counts reported by Sandquist et al. (1996) in their 
Table 7 do not show a dip at the RR Lyrae level---a conclusion
independently supported by Markov's (1997) photometry
(but see also Reid 1996).

Another possible case of HB bimodality is provided by the
relatively metal-rich GC M69 = NGC 6637. As pointed out by
Ferraro et al. (1994; cf. their Sect. 7), a red HB clump seems
to be accompanied by what appears to be a quite long blue tail.
And yet, no RR Lyrae variables have been reported to be cluster 
members, according to the latest edition of the Sawyer Hogg 
(1973) catalogue on variable stars in GCs (Clement 1996).
Likewise, the GC Palomar 2 has been included in Table 4 as
yet another example of HB bimodality, since the CMD of
Harris et al. (1997a) clearly shows both a red clump
and an extended blue tail, but few RR Lyrae variables are 
known (or suspected) to be present in the cluster (Clement
1996; Harris et al. 1997a).

Gap clusters are listed in Table 5, where the columns
have the same meaning as in Table 4.

Racine (1971) lists a few additional candidate gap clusters
besides those in Table 5: 
NGC 4147; M2 = NGC 7089; M12 = NGC 6218; and M22 = NGC 6656. 
However, the latest published CMDs for most of these clusters
(NGC 4147: Auri\`ere \& Lauzeral 1991; M2: Auri\`ere \& Cordoni
1983, Montgomery 1995; M12: Montgomery 1995, Brocato et al. 1996; 
M22: Samus et al. 1995) either do not support
his suggestion or are not suitable for a reliable confirmation.
The presence of a gap on the blue HB of M5 (Brocato,
Castellani, \& Ripepi 1995) is not clearly supported by the
extensive CMD study by Sandquist et al. (1996), although a
gap-like feature is also present in Reid's (1996) photometry
(cf. his Fig. 4).
Similarly, while according to Ferraro, Fusi Pecci, \& Buonanno 
(1992b) NGC 5897 contains a gap on the blue HB, a similar
feature does not seem to be present in the CMD by
Sarajedini (1992). The M55 gap noted by Desidera (1996) is 
not very pronounced, and further analysis of this cluster 
is encouraged. 

Fusi Pecci et al. (1993) report that NGC 1261 and 
M30 = NGC 7099 have HB gaps (cf. their Table 1). However, 
Ferraro et al. (1993) have remarked that ``the blue
HB tail (\ldots) [of NGC 1261]
does not show any marked gap or clump,"
so we do not classify this object as a gap cluster. With 
regard to M30, neither the {\em Hubble Space Telescope} 
(HST) ultraviolet and visual CMDs published by Yanni et al.
(1994) and Mould et al. (1996), nor the ground-based 
photometry by Montgomery (1995), Bergbusch (1996), 
or Burgarella \& Buat (1996), seem to show 
HB gaps either. M3 is also included in the Fusi Pecci 
et al. list of gap clusters. Although this may eventually prove
to be the case (see Ferraro et al. 1997a for a detailed
discussion), we shall not classify M3 as a gap cluster
in the present paper.

The case of NGC 288 deserves special attention, since the 
classic HB gap reported by Buonanno et al. (1984) and Bergbusch 
(1993)---a feature which, according to Crocker et al. (1988), has
a null probability of being a statistical fluctuation---seems 
to have completely vanished in the latest, high-quality
CCD analysis by Kaluzny (1996), although the latter study shows
that a few extremely hot HB stars may be well separated from the
main body of the HB. The famous NGC 288 gap is not evident
in Bolte's (1992) study either, although Bergbusch (1993) not
only confirms it, but also suggests that an additional ``distinct"
gap may be present on the blue HB. In the case of $\omega$Cen,
the reader may find it instructive to inspect Figs. 12 and 
13 in Kaluzny et al. (1996), along with Figs. 8 and 9 in
Kaluzny et al. (1997), to evaluate the degree to which the
appearance of gap-like features may depend on the specific
fields analyzed, sample sizes, quality of
the data, and other observational details. Similarly, we draw 
attention to Figs. 1 and 2 in Kaluzny (1997), where CMDs for
NGC 6397 are presented. While a gap is quite evident in his 
Fig. 1, it is absent in his Fig. 2, where different reduction 
procedures and data selection and treatment were adopted. 
A blue-HB gap was also not found in the previous study of 
this cluster by Alcaino et al. (1987).

M70 = NGC 6681 is included in Table 5
because the Watson et al. (1994) UV-visible CMD
shows a gap (not noted by those authors) at 
$T_{\rm eff} \approx 8,\!700 \, {\rm K}$ (cf. their Fig. 2).
The gap is not clear in the ground-based optical photometry 
by Brocato et al. (1996), though.

Worth noting is the fact that,
in general, gaps have been reported
mostly on the blue HB. Fusi Pecci et al. (1992) discuss,
however, the possible existence of gaps on the red HB 
(in particular in the case of M3; see also Renzini \& Fusi
Pecci 1988), which they argue may represent the progeny of 
blue straggler stars. As far as we are aware, the 
NGC 6638 and IC 4499 gaps are the only ones 
which have been reported to lie {\em inside} 
the instability strip, although we note, in passing, that
Fig. 10 in Buonanno et al. (1994) for M3 also shows evidence
for a similar gap. Of course, the uncertainties involved
in the analysis of this kind of gap are very large, since
determining equilibrium colors for RR Lyrae variables is a far
from trivial task. Reid (1996), for instance, notices
that a gap between the RR Lyrae strip and the red HB may or may
not exist in M5, depending on which technique one employs.

Analysis of Table 4 discloses that HB bimodality only occurs
in intermediate-to-high metallicity GCs. No correlation
with central concentration seems to be present. Table 5
fails to reveal any correlation between the presence of
HB gaps and either [Fe/H] or core concentration. However,
since it seems likely that some of the detected gaps 
may be due to statistical fluctuations (cf. Sec. 5), any
real correlations would tend to be erased by the ``noise"
introduced by these ``spurious" detections.

\clearpage

\begin{figure}[t]
 \plotone{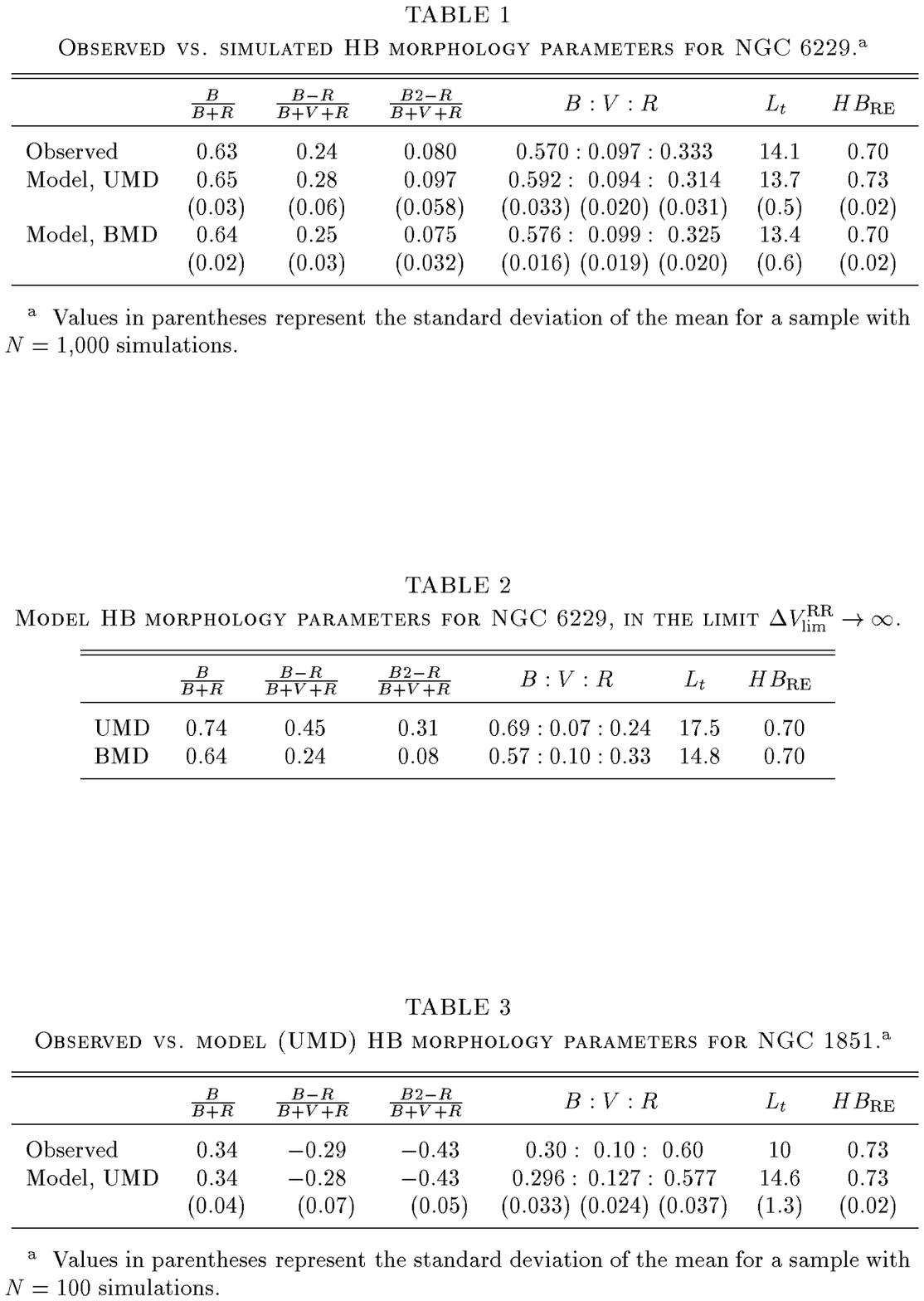}
\end{figure}

\clearpage

\begin{figure}[t]
 \plotone{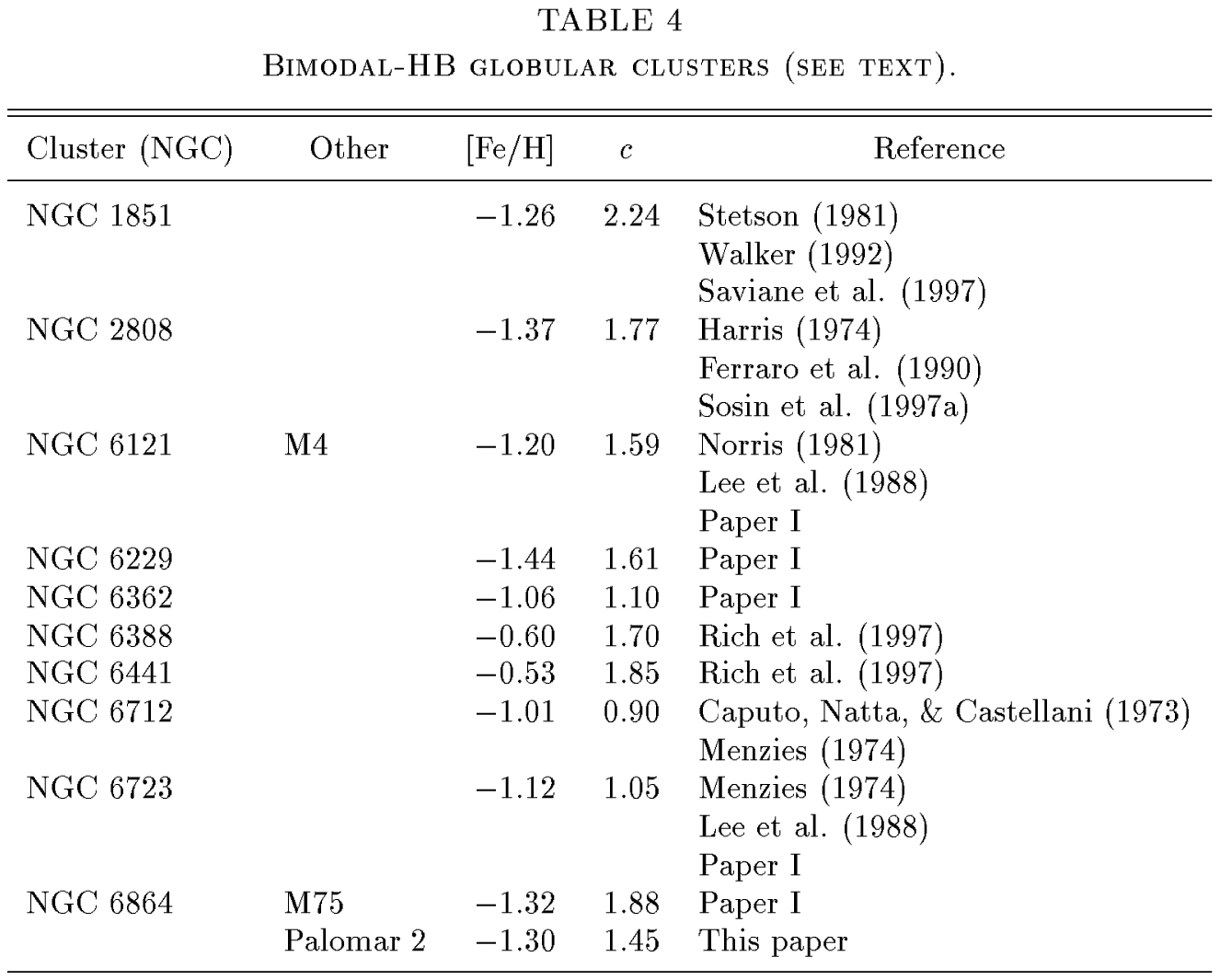}
\end{figure}

\clearpage

\begin{figure}[t] 
 \plotone{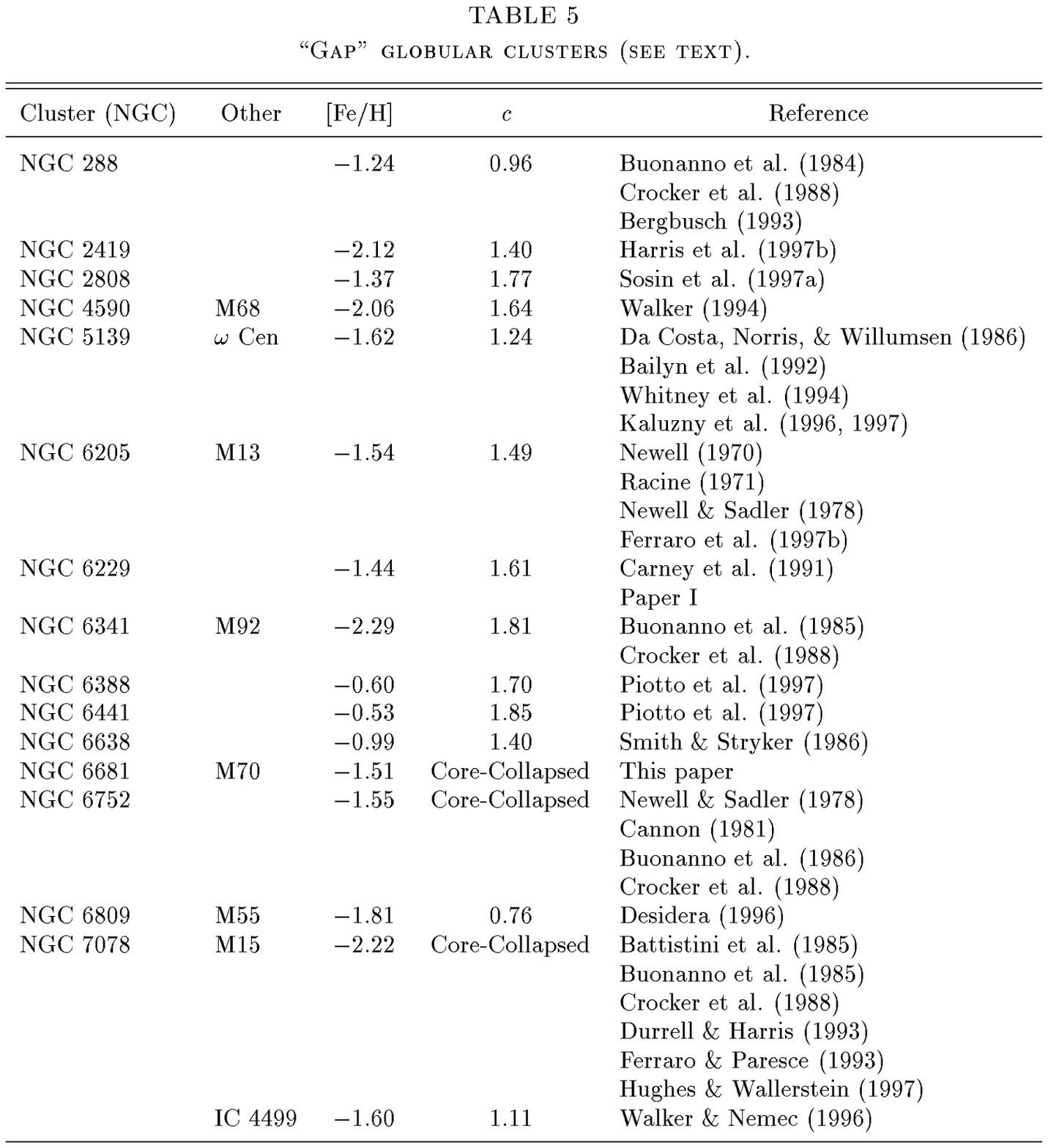}
\end{figure}

\newpage

%
%                                                One column figure
%----------------------------------------------------------- Fig. 1
\begin{figure}[t] 
 \figurenum{1}
 \plotone{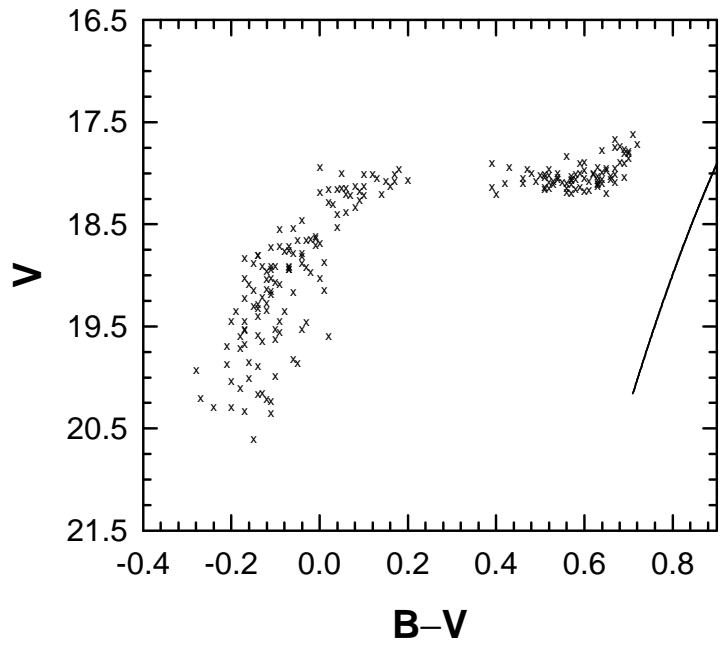}
 \caption{The HB region of the NGC 6229 CMD from Paper I.
          RR Lyrae variables have not been plotted. The position of
          the RGB is schematically indicated as a solid line.
         }
\end{figure}

\clearpage

%
%                                                One column figure
%----------------------------------------------------------- Fig. 2
\begin{figure}[t] 
 \figurenum{2}
 \plotone{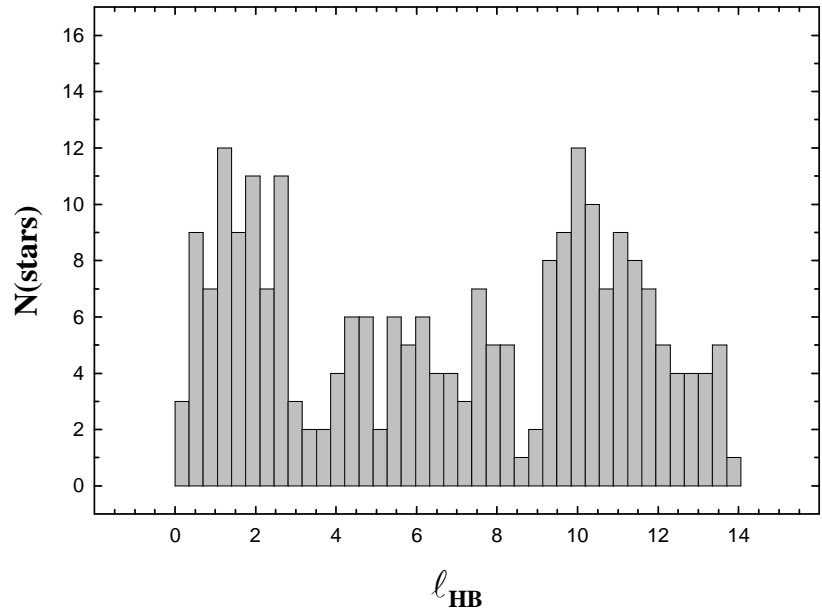}
 \caption{$\ell_{\rm HB}$ distribution for NGC 6229 from the data
          in Fig. 1. A total of 31 RR Lyrae stars was included
          with randomly distributed colors inside the instability
          strip.
         }
\end{figure}

\clearpage

%
%                                                One column figure
%----------------------------------------------------------- Fig. 3
\begin{figure}[t] 
 \figurenum{3}
 \plotone{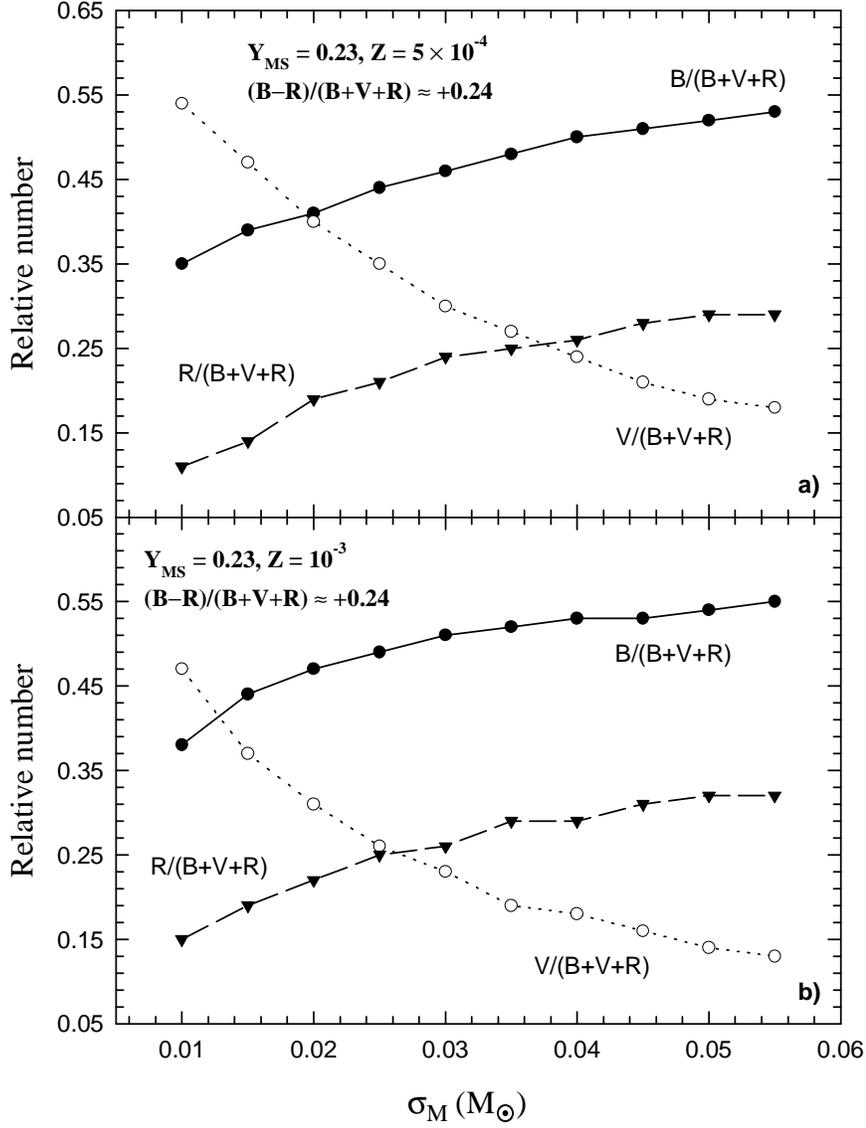}
 \caption{The variation of $V/(B+V+R)$ as a function of
          the HB mass dispersion $\sigma_M$ as compared 
          with the corresponding variations in $B/(B+V+R)$ and
          $R/(B+V+R)$ for two metallicities,
          $Z = 0.0005$ (panel a) and $Z = 0.001$ (panel b).
          The Lee-Zinn parameter $(B-R)/(B+V+R)$
          was held fixed at the value appropriate for NGC 6229.
         }
\end{figure}

\clearpage

%
%                                                One column figure
%----------------------------------------------------------- Fig. 4
\begin{figure}[t] 
 \figurenum{4}
 \plotone{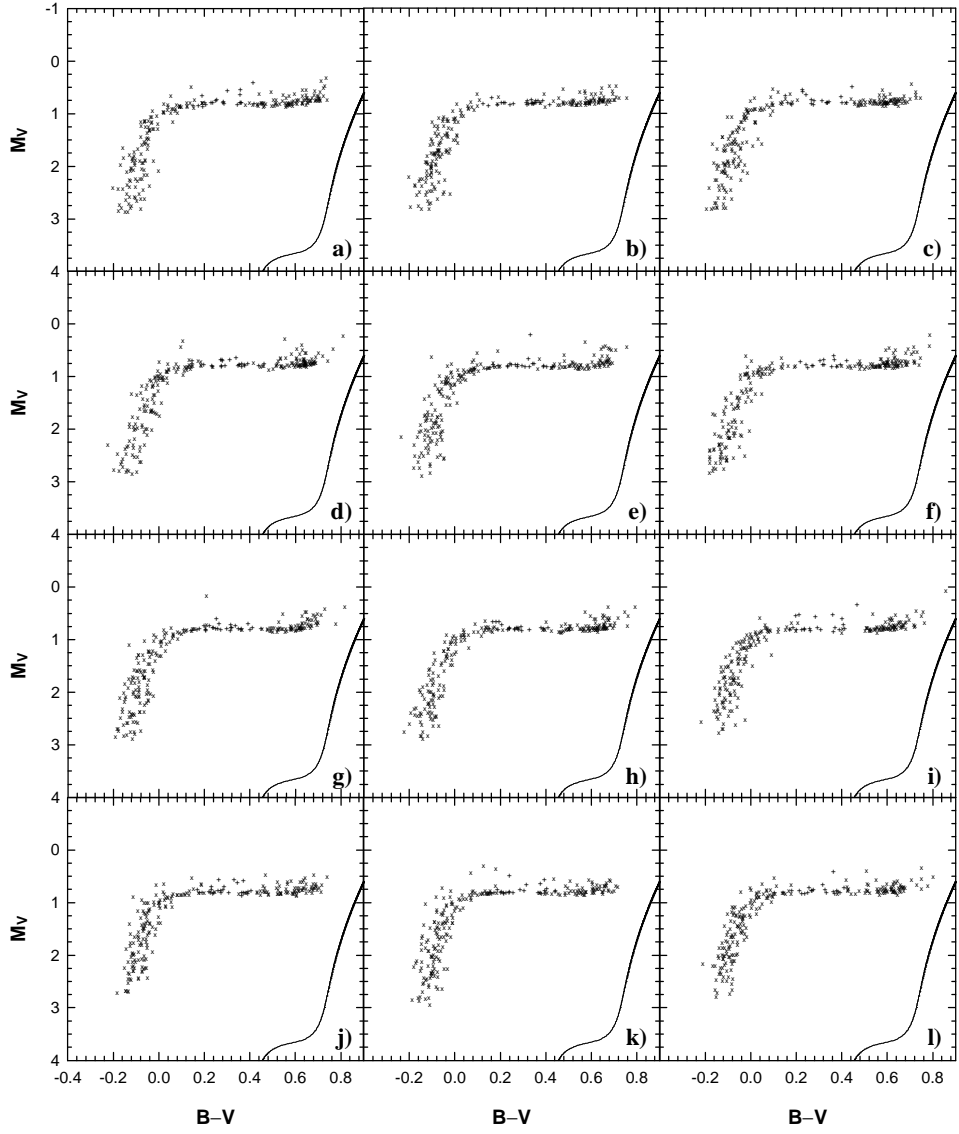}
 \caption{Random sample of 12 HB simulations for NGC 6229, obtained 
          from the set of 1,000 synthetic HBs computed for the unimodal
          mass distribution case discussed in Sec. 4. RR Lyrae variables 
          are indicated by plus signs.
          The solid lines show the position of an RGB
          evolutionary track with $M = 0.820\, M_{\sun}$.
         }
\end{figure}

\clearpage

%
%                                                One column figure
%----------------------------------------------------------- Fig. 5
\begin{figure}[t] 
 \figurenum{5}
 \plotone{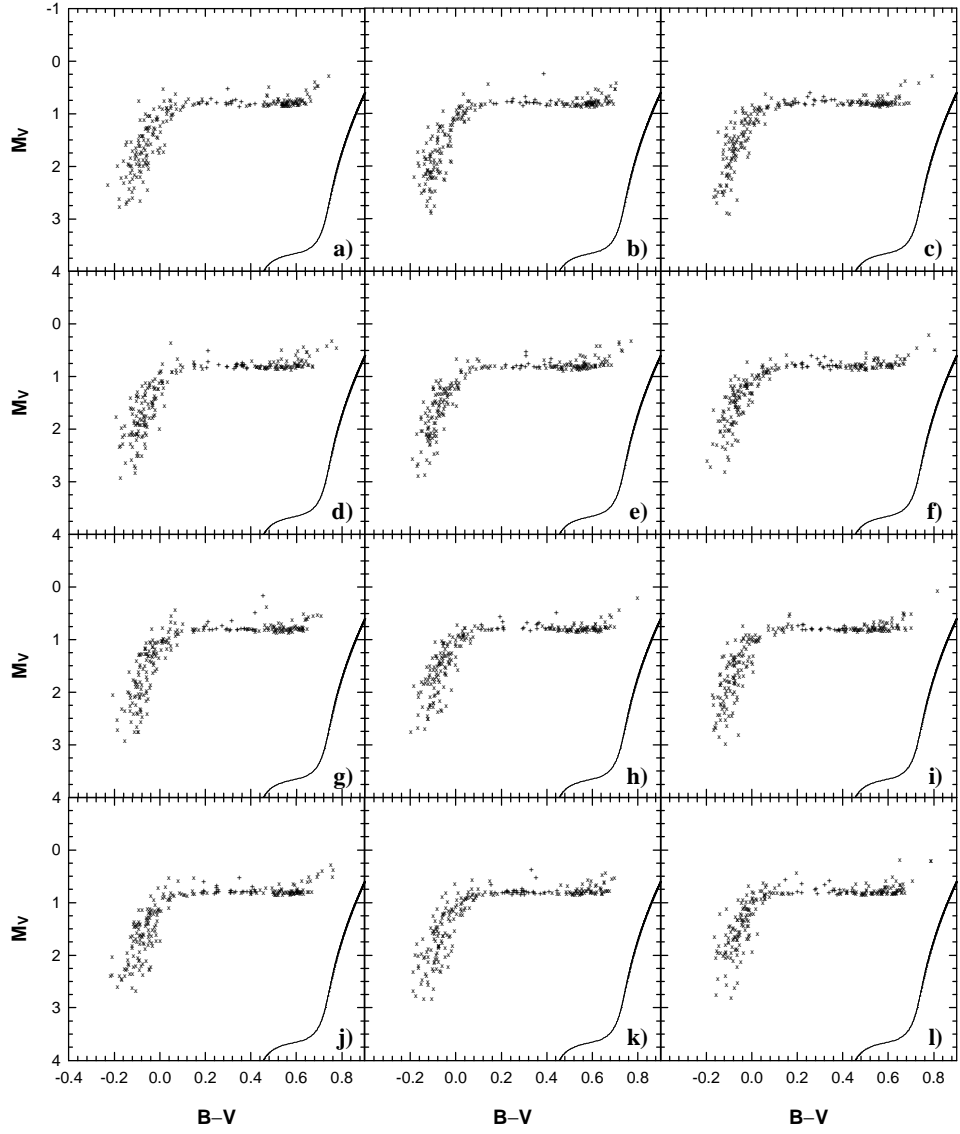}
 \caption{As in Fig. 4, but for the bimodal mass distribution case
          discussed in Sec. 4.
         }
\end{figure}

\clearpage

%
%                                                One column figure
%----------------------------------------------------------- Fig. 6
\begin{figure}[t] 
 \figurenum{6}
 \plotone{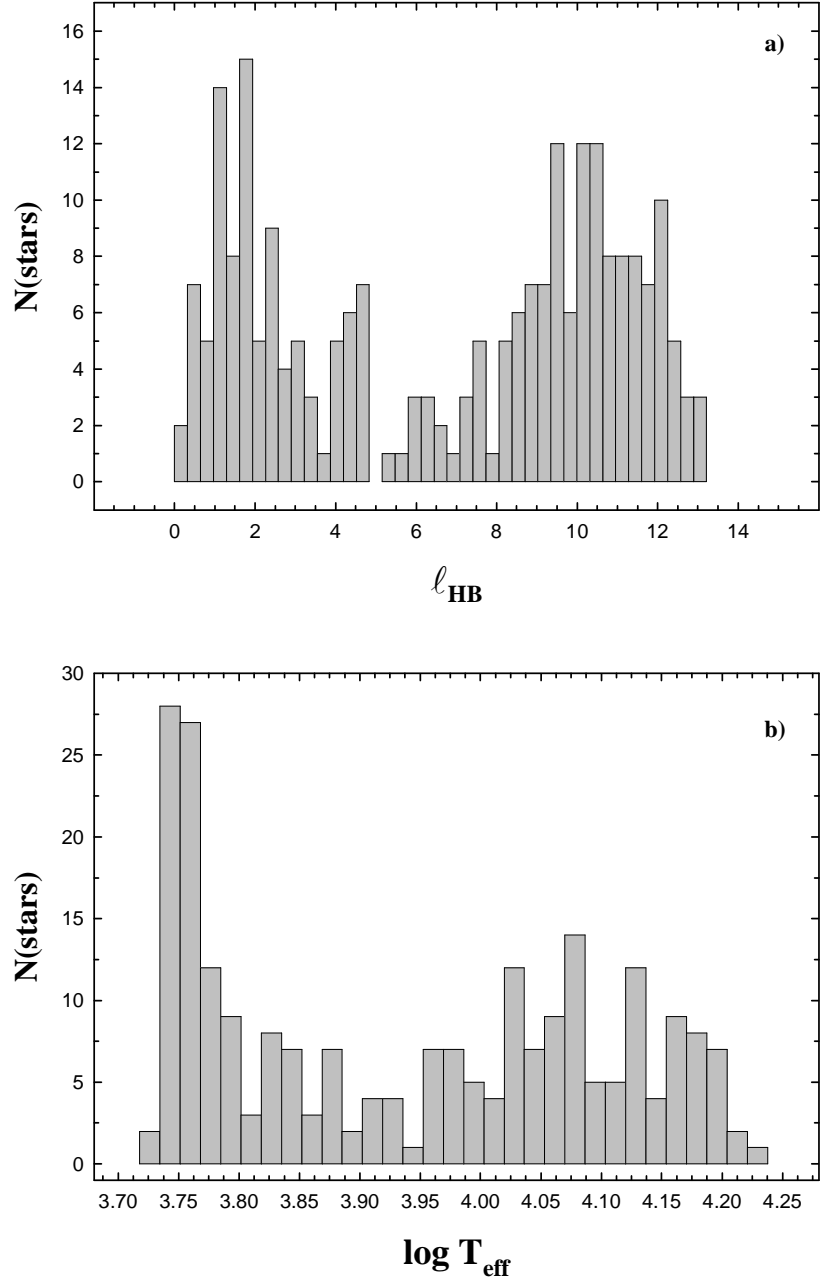}
 \caption{Histograms showing the $\ell_{\rm HB}$ (panel a) and
          $\log\,T_{\rm eff}$ (panel b) distribution for the
          HB simulation (unimodal mass distribution) 
          displayed in Fig. 4b.
         }
\end{figure}

\clearpage

%
%                                                One column figure
%----------------------------------------------------------- Fig. 7
\begin{figure}[t] 
 \figurenum{7}
 \plotone{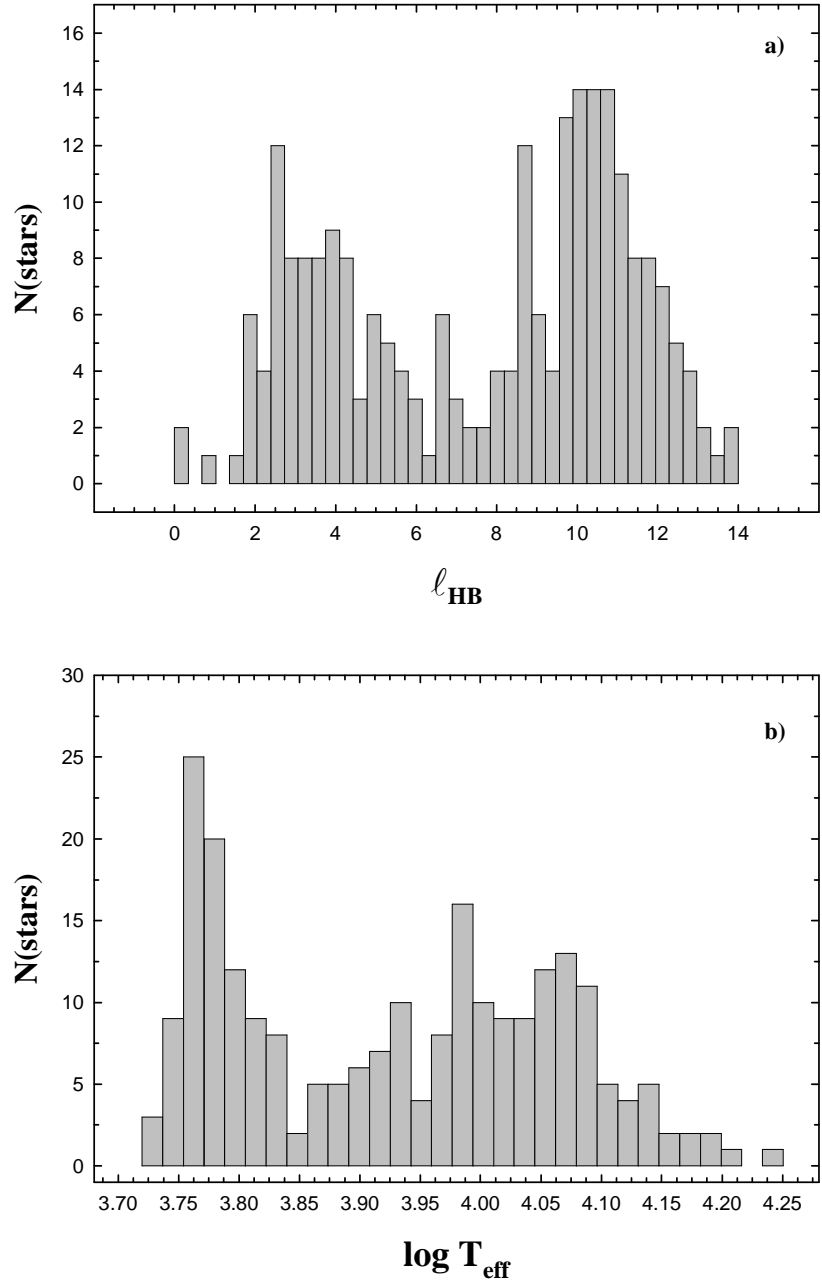}
 \caption{As in Fig. 6, but for the HB simulation
          (bimodal mass distribution) displayed in Fig. 5f.
         }
\end{figure}

\clearpage

%
%                                                One column figure
%----------------------------------------------------------- Fig. 8
\begin{figure}[t] 
 \figurenum{8}
 \plotone{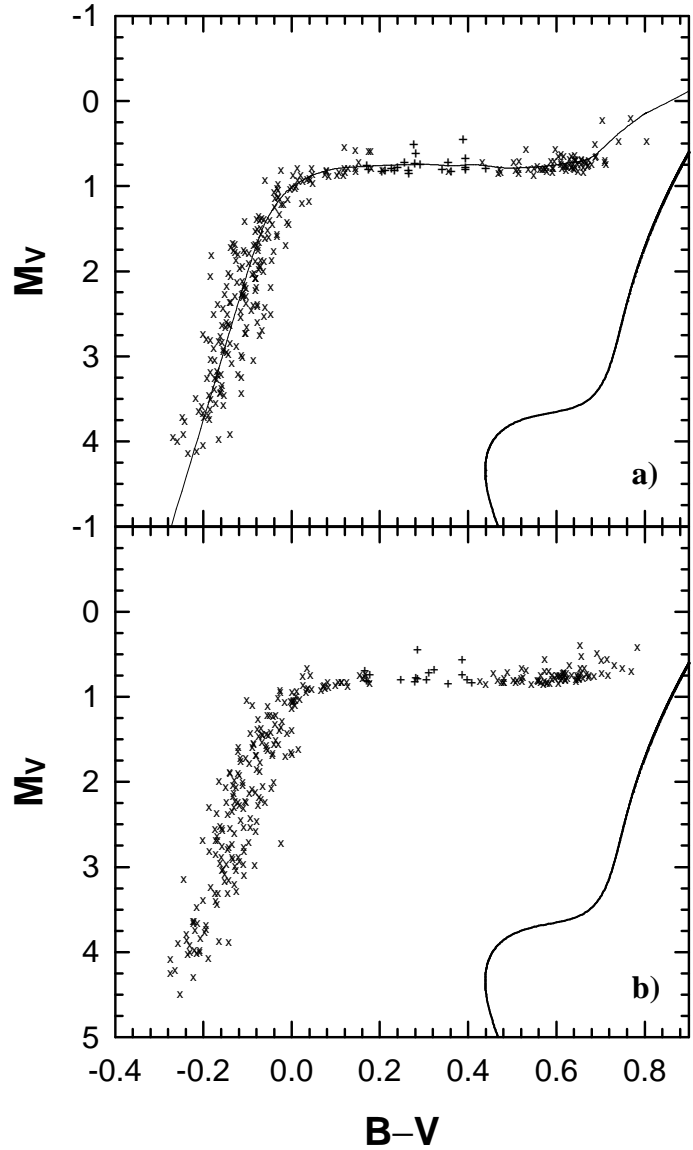}
 \caption{Two HB simulations for the same
          unimodal mass distribution employed in
          generating Fig. 4, but without 
          truncating the distribution at the faint limit.
          The computed HB ridgeline is shown as a thin solid line
          in panel a), and the RGB locus for a $M = 0.820\, M_{\sun}$
          star is given as a thick solid line in both panels.
         }
\end{figure}

\clearpage

%
%                                                One column figure
%----------------------------------------------------------- Fig. 9
\begin{figure}[t] 
 \figurenum{9}
 \plotone{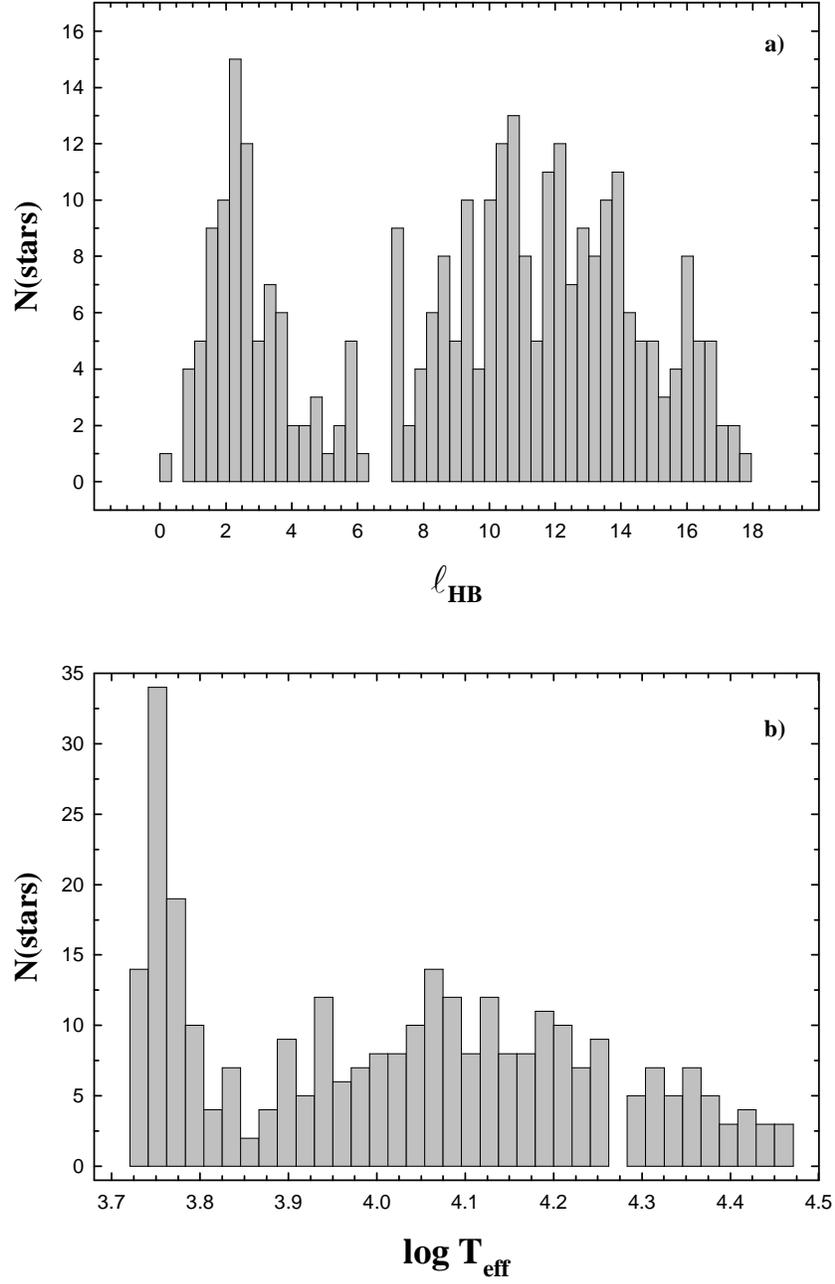}
 \caption{Histograms showing the $\ell_{\rm HB}$ (panel a) and
          $\log\,T_{\rm eff}$ (panel b) distributions for the
          HB simulation displayed in Fig. 8b.
         }
\end{figure}

\clearpage

%
%                                                One column figure
%----------------------------------------------------------- Fig. 10
\begin{figure}[t] 
 \figurenum{10}
 \plotone{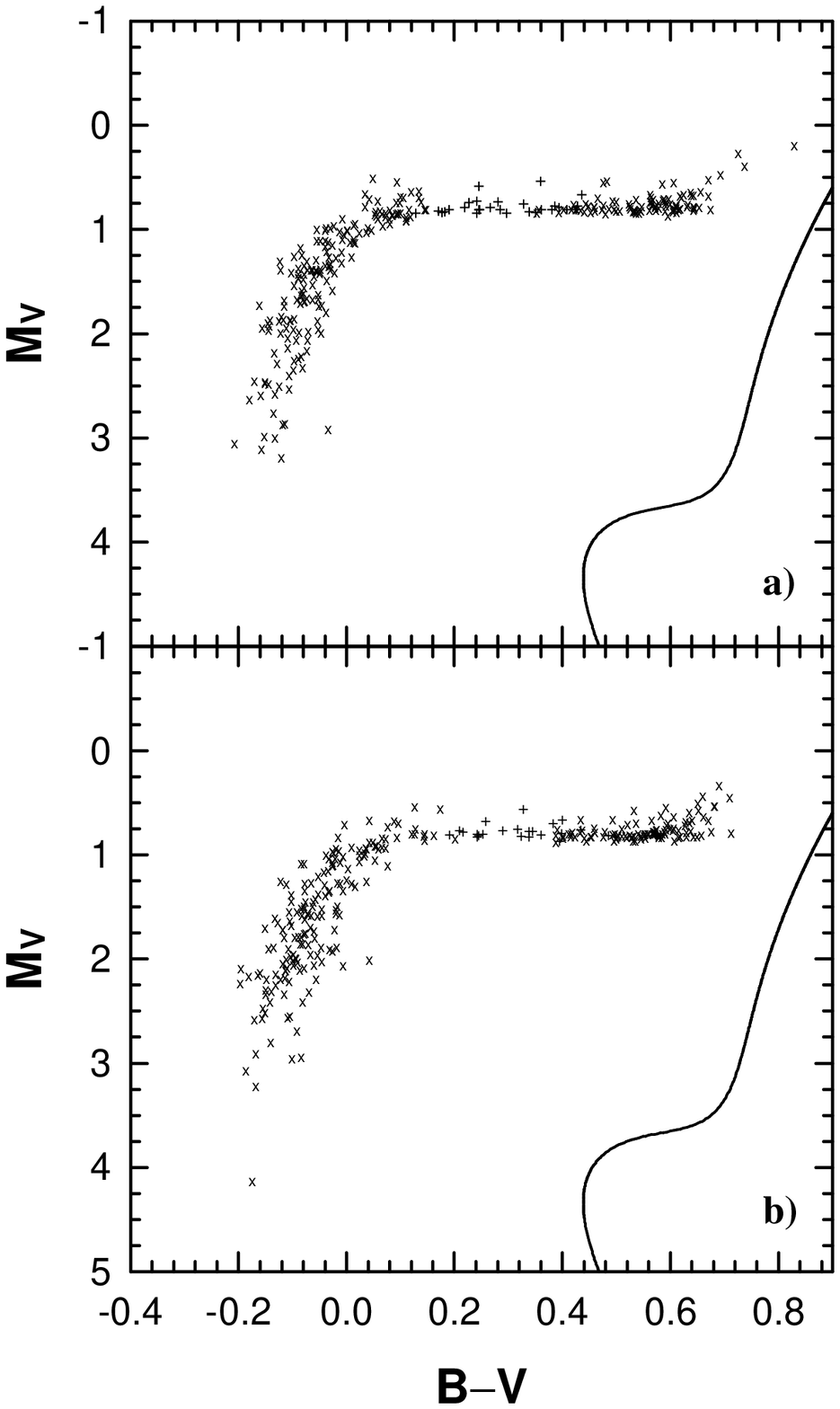}
 \caption{Same as Fig. 8, but for a bimodal mass distribution.
         }
\end{figure}

\clearpage

%
%                                                One column figure
%----------------------------------------------------------- Fig. 11
\begin{figure}[t] 
 \figurenum{11}
 \plotone{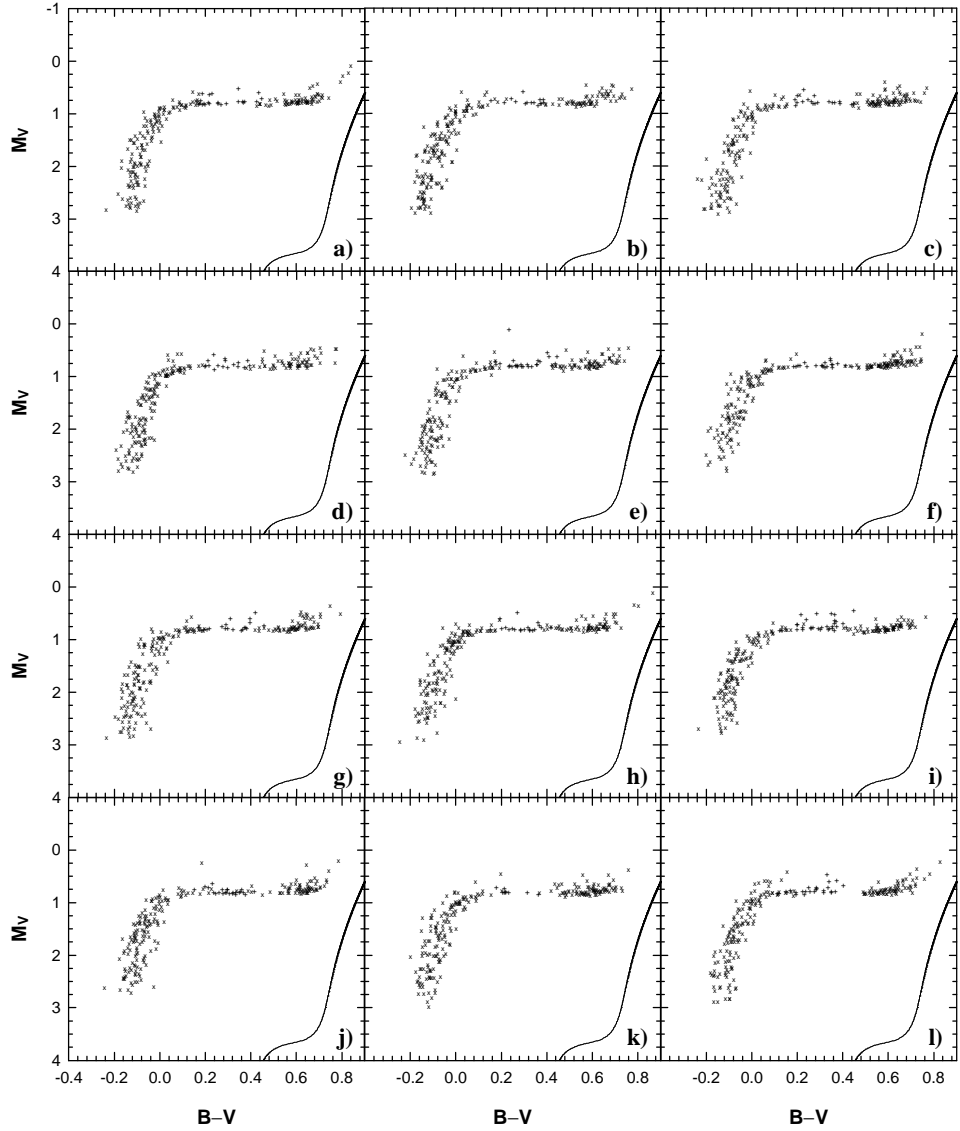}
 \caption{Same as Fig. 4, but for a sequence of synthetic
          HBs showing pronounced gaps at increasingly redder colors
          [going from a) to l)].
         }
\end{figure}

\clearpage

%
%                                                One column figure
%----------------------------------------------------------- Fig. 12
\begin{figure}[t] 
 \figurenum{12}
 \plotone{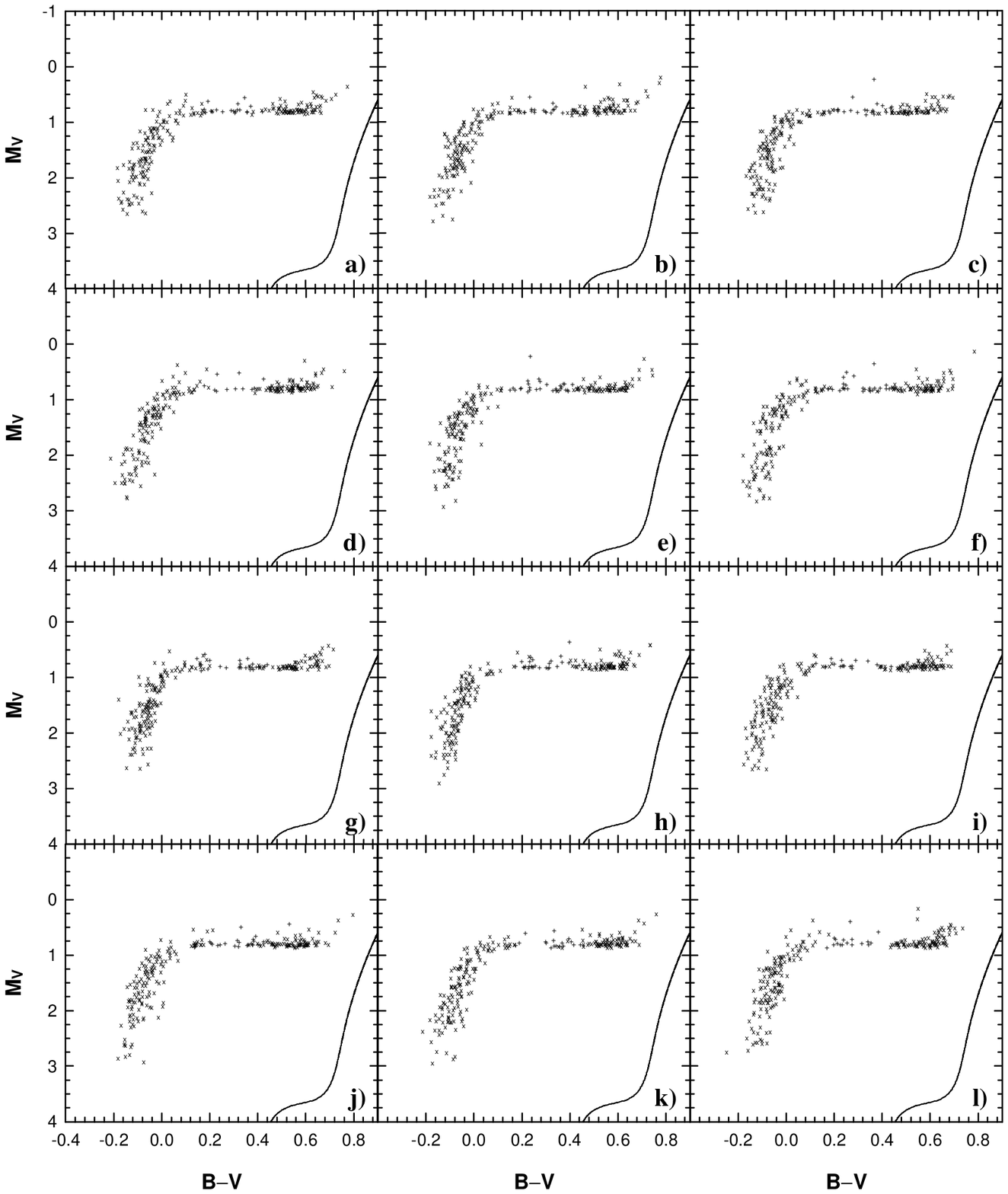}
 \caption{Same as Fig. 11, but for a bimodal mass distribution.
         }
\end{figure}

\clearpage

%
%                                                One column figure
%----------------------------------------------------------- Fig. 13
\begin{figure}[t] 
 \figurenum{13}
 \plotone{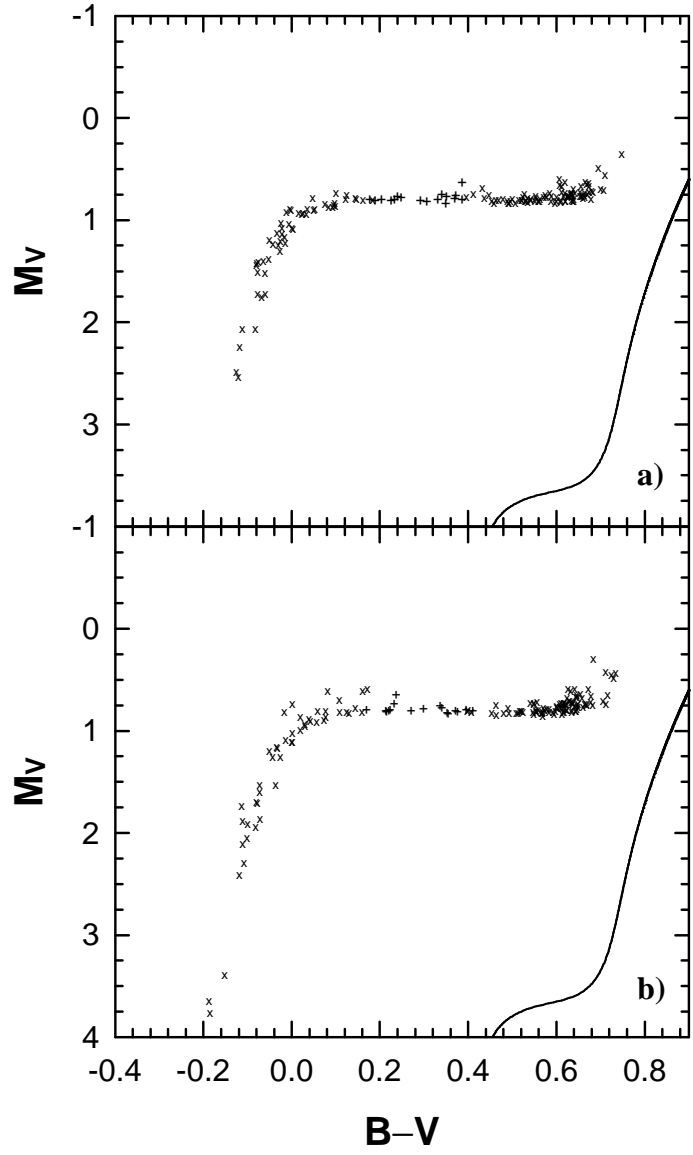}
 \caption{Examples of synthetic HBs for NGC 1851. 
         }
\end{figure}

\clearpage

%
%                                                One column figure
%----------------------------------------------------------- Fig. 14
\begin{figure}[t] 
 \figurenum{14}
 \plotone{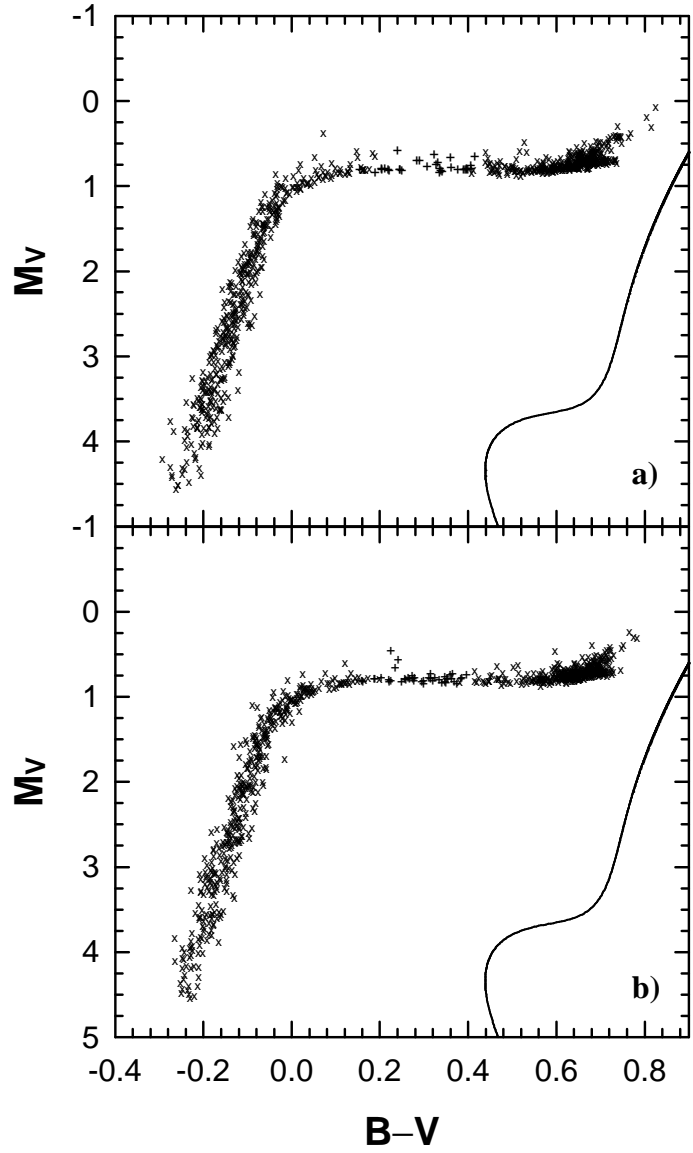}
 \caption{Examples of synthetic HBs with a unimodal mass
          distribution for NGC 2808. These 
          simulations fail to provide a
          satisfactory match to the observed CMD.
         }
\end{figure}

\clearpage

%
%                                                One column figure
%----------------------------------------------------------- Fig. 15a
\begin{figure}[t] 
  \epsscale{0.80}
  \plotone{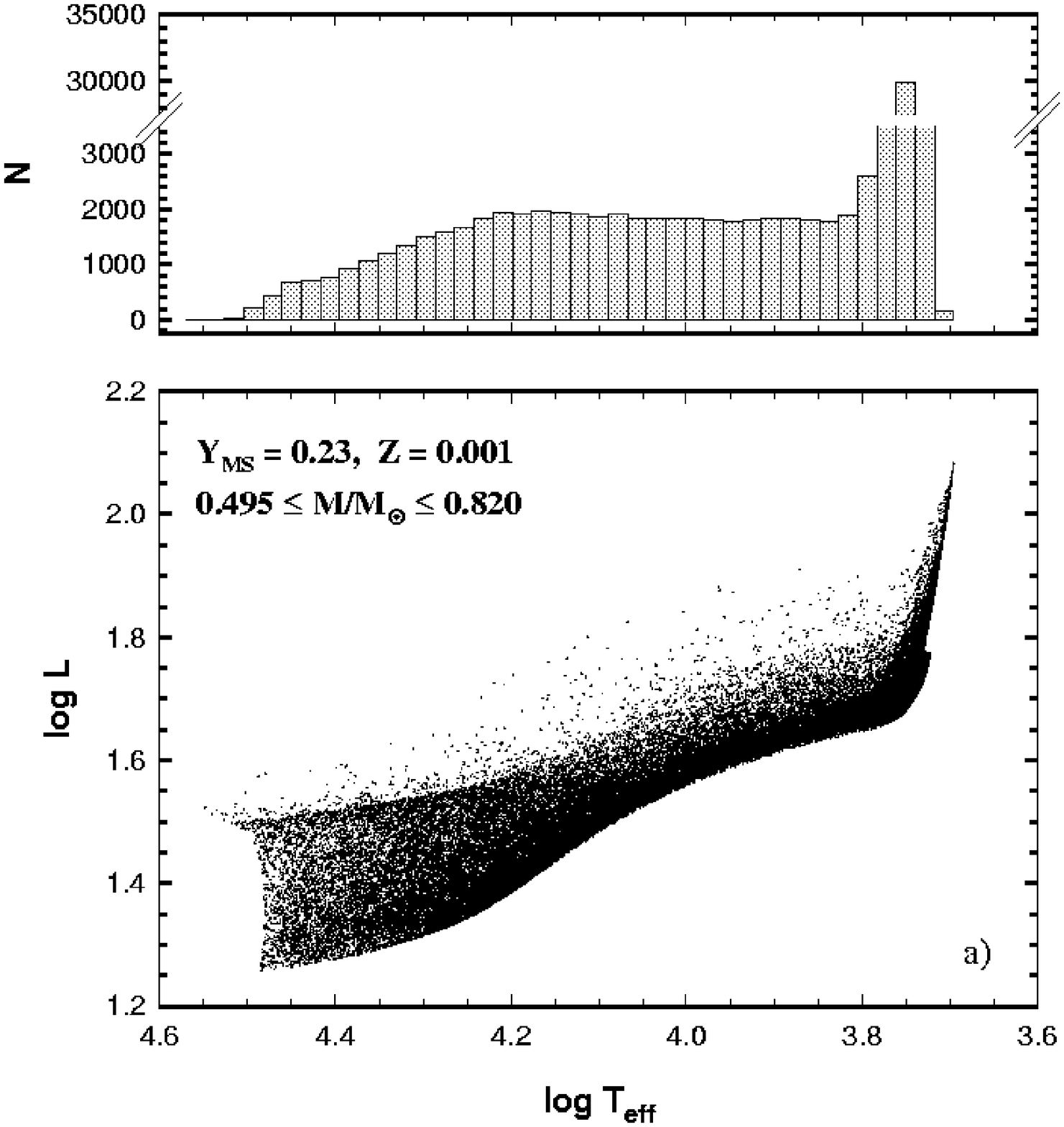}
\end{figure}

\clearpage

%
%                                                One column figure
%----------------------------------------------------------- Fig. 15b
\begin{figure}[t] 
 \figurenum{15}
 \epsscale{0.80}
 \plotone{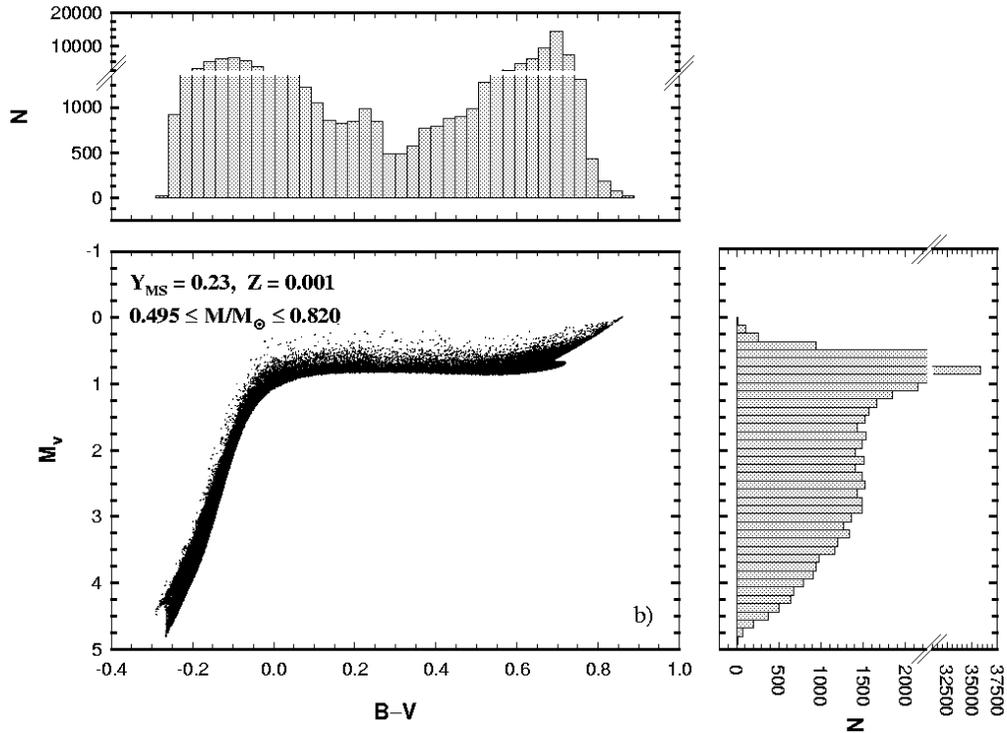}
 \caption{Reference synthetic HB simulation with 100,000 stars, 
          computed for a uniform ZAHB mass
          distribution in the indicated mass range. Panel a) shows
          the distribution in the ($\log\,L$, $\log\,T_{\rm eff}$)
          plane, and panel b) the distribution in the
          ($M_V$, $\bv$) plane. Histograms giving the corresponding
          distributions in $\log\,T_{\rm eff}$, \bv, and $M_V$ are
          also given.}
\end{figure}

\clearpage

%
%                                                One column figure
%----------------------------------------------------------- Fig. 16
\begin{figure}[t] 
 \figurenum{16}
 \plotone{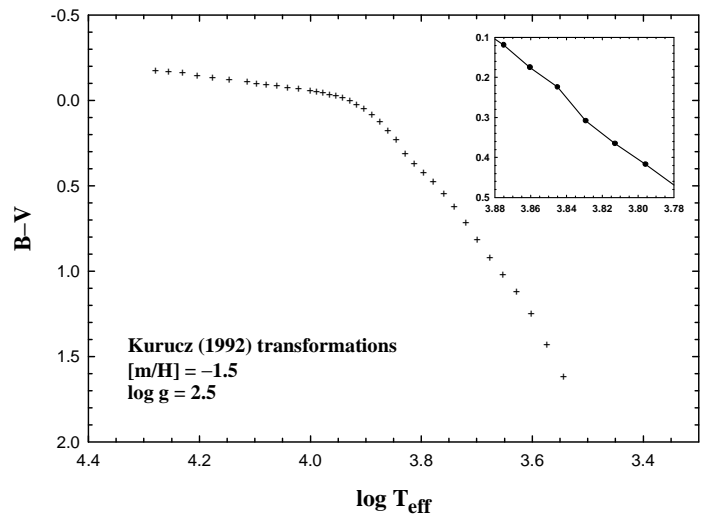}
 \caption{Color--temperature relation from the Kurucz (1992) model 
          atmospheres, for the gravity and metallicity 
          values shown on the lower left-hand corner.
         }
\end{figure}

\clearpage

%
%                                                One column figure
%----------------------------------------------------------- Fig. 17
\begin{figure}[t] 
 \figurenum{17}
 \plotone{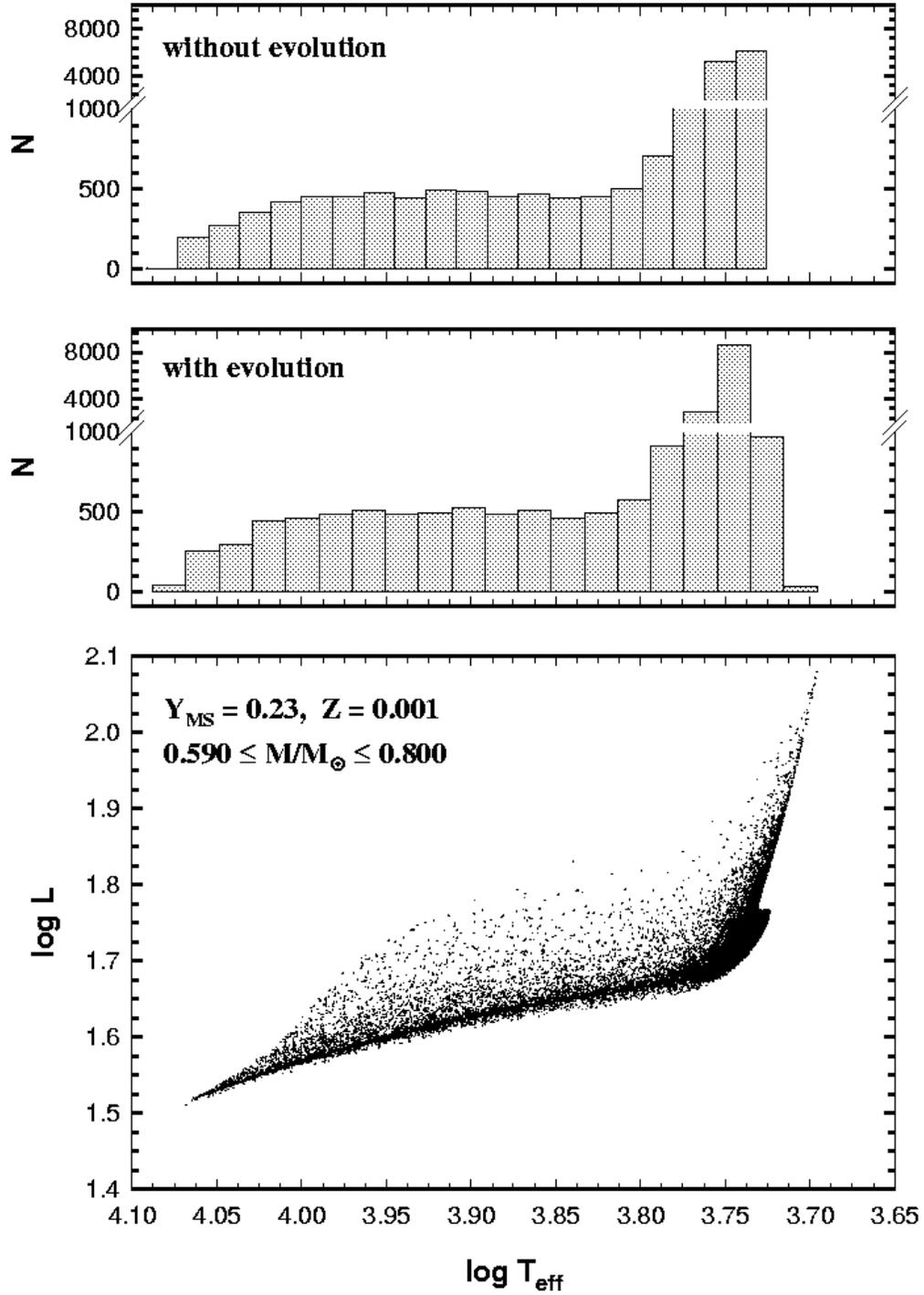}
 \caption{A synthetic HB with 20,000 stars, 
          computed for a uniform ZAHB mass
          distribution in the indicated mass range, 
          is shown in the ($\log\,L$, $\log\,T_{\rm eff}$)
          plane (bottom panel). The mass range was chosen so as
          to mimic Walker's (1992) Fig. 8 as closely as possible.
          Histograms for $T_{\rm eff}$ are given 
          both for the case which includes evolution away from
          the ZAHB (middle panel) and the case where evolution
          away from the ZAHB is suppressed (upper panel).
         }
\end{figure}

\clearpage

%
%                                                One column figure
%----------------------------------------------------------- Fig. 18
\begin{figure}[t] 
 \figurenum{18}
 \plotone{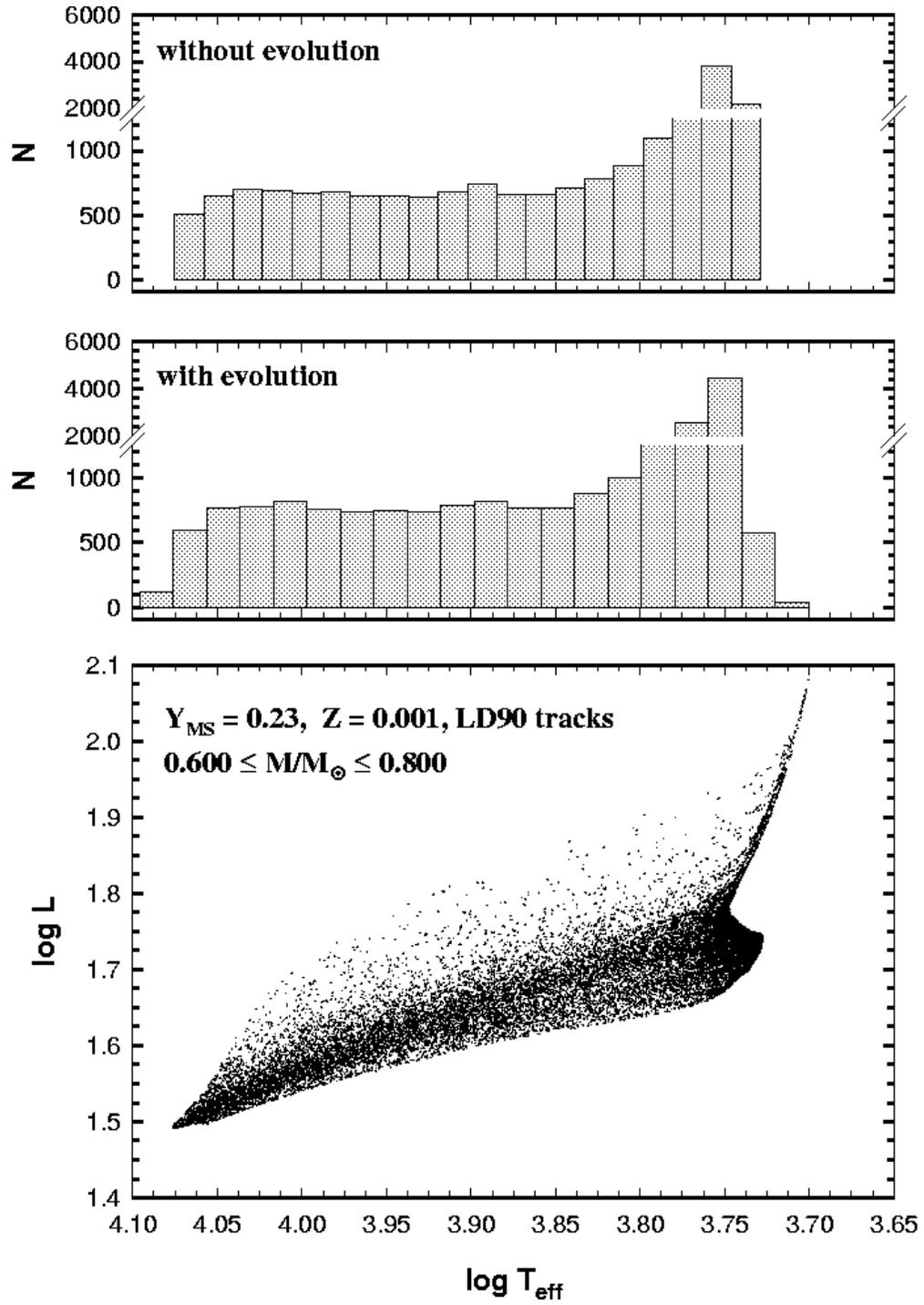}
 \caption{As in Fig. 17, but using the Lee \& Demarque (1990, LD90)
          evolutionary tracks.}
\end{figure}

\end{document}